# A Tight VC-Dimension Analysis of Clustering Coresets with Applications

Vincent Cohen-Addad  Andrew Draganov  Matteo Russo  David Saulpic
Chris Schwiegelshohn

**Abstract**

We consider coresets for $k$-clustering problems, where the goal is to assign points to centers minimizing powers of distances. A popular example is the $k$-median objective $\sum_p \min_{c \in C} dist(p, C)$. Given a point set $P$, a coreset $\Omega$ is a small weighted subset that approximates the cost of $P$ for all candidate solutions $C$ up to a $(1 \pm \varepsilon)$ multiplicative factor. In this paper, we give a sharp VC-dimension based analysis for coreset construction. As a consequence, we obtain improved $k$-median coreset bounds for the following metrics:

- Coresets of size $\tilde{O}\left(k\varepsilon^{-2}\right)$ for shortest path metrics in planar graphs, improving over the bounds $\tilde{O}\left(k\varepsilon^{-6}\right)$ by [Cohen-Addad, Saulpic, Schwiegelshohn, STOC'21] and $\tilde{O}\left(k^2\varepsilon^{-4}\right)$ by [Braverman, Jiang, Krauthgamer, Wu, SODA'21].
- Coresets of size $\tilde{O}\left(kd\ell\varepsilon^{-2}\log m\right)$ for clustering $d$-dimensional polygonal curves of length at most $m$ with curves of length at most $\ell$ with respect to Frechet metrics, improving over the bounds $\tilde{O}\left(k^3 d\ell\varepsilon^{-3}\log m\right)$ by [Braverman, Cohen-Addad, Jiang, Krauthgamer, Schwiegelshohn, Toftrup, and Wu, FOCS'22] and $\tilde{O}\left(k^2 d\ell\varepsilon^{-2}\log m \log |P|\right)$ by [Conradi, Kolbe, Psarros, Rohde, SoCG'24].

## 1 Introduction

Coresets are a staple of both learning theory and big data algorithms. Given a set of points $P$ and a set of queries $\mathcal{Q}$ along with an associated loss function $f : P \times \mathcal{Q} \to \mathbb{R}_{\geq 0}$, a coreset of $P$ is a (typically weighted) subset $\Omega$ of $P$ with the property that $f(\Omega, Q) \approx f(P, Q)$ for all $Q \in \mathcal{Q}$.

Coresets have found numerous applications since their introduction in the early 2000s. One typical use case consists of clustering problems, such as $k$-median. Given a set $P$ of $n$ (possibly weighted) points in a metric space, the $(k, z)$-clustering objective asks to find a set $\mathcal{S}$ of $k$ *center* points so as to minimize the sum of the distances raised to the $z$th power from each point in $P$ to its closest representative:

$$f(P, \mathcal{S}) = \text{cost}(P, \mathcal{S}) := \sum_{p \in P} w_p \cdot \min_{c \in \mathcal{S}} \text{dist}^z(p, c),$$

where $w_p$ denotes a non-negative weight and dist denotes the distance function associated with the underlying metric. The for us most important example is the $k$-median objective, where $z = 1$. In this case, we aim to find a coreset $\Omega$ such that for all candidate solutions $S$, i.e., set of $k$ centers

$$(1 - \varepsilon) \cdot \text{cost}(P, \mathcal{S}) \leq \text{cost}(\Omega, \mathcal{S}) \leq (1 + \varepsilon) \cdot \text{cost}(P, \mathcal{S}).$$

For many important metric spaces, including Euclidean spaces, doubling metrics, and finite metrics, the question of obtaining optimal coreset sizes has been effectively settled. The state of the art coreset sizes are $\tilde{O}(k\varepsilon^{-2} \cdot \min(\sqrt[3]{k}, \varepsilon^{-1}, d))$ [5, 20, 39] for $d$-dimensional Euclidean spaces, $\tilde{O}(kD\varepsilon^{-2})$ [18, 39] for doubling spaces with doubling constant $D$, and $\tilde{O}(k\varepsilon^{-2}\log n)$ for finite $n$ point metrics [18, 21], all of which are tight up to logarithmic factors in $k$ or $\varepsilon$.[1]

In recent years, newfound insights from learning theory towards constructing good, unbiased estimators for these cost functions have fueled research on coresets. To prove that an unbiased estimator yields a coreset, the analysis has to (i) show that the variance of the estimator is small enough so as to allow to (ii) take a union bound over all solutions. The variance bounds for various estimators are independent of the underlying metric space and are at this stage of coreset research very well understood. Our ability to find small enumerators of the solution space, which are often referred to as *nets* is still an open problem for numerous metrics.

[1] We use $\tilde{O}(x)$ to suppress polylog$(x)$ factors.

One of the most important complexity measures for designing nets in learning theory is the VC dimension. For clustering, the VC dimension determines the maximum number of dichotomies that we can induce by placing a center $c$ with some flexible radius $r$ and labeling the points with distance at least $r$ to $c$ by 1 and the points with distance less than $r$ to $c$ by 0. Specifically, the size $d_{\mathrm{VC}}$ of the largest set $P \subseteq X$ for which we can generate all possible dichotomies is the VC dimension. The first sampling-based coreset results heavily relied on nets derived by way of the VC-dimension [15, 40, 28, 30, 36]. Indeed, to this day, there exist several metrics for which the VC dimension is the only known technique towards constructing coresets. Examples include shortest path metrics in minor free graphs [10, 11] and Frechet and Hausdorff metrics [11, 13, 24].

Perhaps surprisingly due to its well studied nature in the field of coresets, our ability to construct coresets via the VC dimension is not fully understood. Most constructions reduce the problem to some variant of uniform sampling in cases where we only have bounds for the unweighted VC dimension (see for example [8] for shortest path metrics in planar graphs and [26] for Frechet metrics). Unfortunately, these reductions [15, 8] always incur a loss in parameters because they require reducing the instance to a setting where uniform sampling becomes viable. Alternative methods for non-uniform sampling [44, 30] likewise incur a loss in parameters. In particular, all constructions have at least a quadratic dependency in the, arguably most important, parameter $k$. This comes in addition to extra dependencies on other parameters, which vary based on the construction method used.

## 1.1 Our Results

Motivated by the current state of the art, we aim towards obtaining a sharp VC-dimension-based analysis of sampling based coresets. Given a point set $P$ in some metric space $(X, \mathrm{dist})$, we let $R(c,r) := \{p \in P \mid \mathrm{dist}(p,c) \geq r\}$. Then we call $\mathcal{R} = \cup_{c,r} R(c,r)$ the range space induced by metric balls. The VC dimension of $\mathcal{R}$ is the size of the largest subset $Y$ of $X$ such that the power set of $Y$ is contained in $\mathcal{R}$.

**Theorem 1.1.** *Let $P$ be a point set in some metric space $(X, \mathrm{dist})$ and let $d_{\mathrm{VC}}$ be the VC dimension of metric balls. Then there exists an $\varepsilon$-coreset for $k$-median of size*

$$\tilde{O}\left(k \cdot d_{\mathrm{VC}} \cdot \varepsilon^{-2}\right).$$

This achieves the first VC-dimension-based coreset bound with a $k \cdot \mathrm{polylog}(k)$ dependency, which is optimal up to polylog factors. We also note that the analysis is tight with respect to the other parameters, again up to polylog factors, in the sense that there exist metric spaces for which the stated bound cannot be improved by any coreset algorithm. For example, in finite $n$ point metrics, this bound recovers the optimal $\tilde{O}\left(k \cdot \log n \cdot \varepsilon^{-2}\right)$ coreset size of [18, 39].[2] Nevertheless we emphasize that the VC dimension may not characterize the optimal coreset size for every metric. For example, in $d$-dimensional Euclidean space, the VC-dimension of the range space induced by Euclidean balls is $d+1$ and thus the VC-dimension based analysis only achieves a $\tilde{O}(k \cdot d \cdot \varepsilon^{-2})$ coreset size, while the bounds mentioned above for the same problem are independent of $d$.

As an immediate consequence, this improves coreset bounds for several metric spaces. Specifically, for shortest path metrics in planar graphs, we obtain $k$-median coresets of size $\tilde{O}\left(k \cdot \varepsilon^{-2}\right)$, improving over the $\tilde{O}\left(k^2 \cdot \varepsilon^{-4}\right)$ bound of [10] and the $\tilde{O}\left(k \cdot \varepsilon^{-6}\right)$ bound of [18]. The full set of applications for all considered metric spaces, including coresets for minor free graphs, Frechet metrics and the Hausdorff metric, is given in Section 5.

For general $(k,z)$ clustering objectives, we obtain the following bounds:

**Theorem 1.2.** *Let $P$ be a point set in some metric space $(X, \mathrm{dist})$ and let $d_{\mathrm{VC}}$ be the VC dimension of metric balls. Then there exists an $\varepsilon$-coreset for $(k,z)$-clustering of size*

$$\tilde{O}\left(k \cdot d_{\mathrm{VC}} \cdot \varepsilon^{-2} \cdot \min(\varepsilon^{-z+1}, k)\right).$$

[2] The VC dimension can only be of the order $\log n$, as there are at most $n$ many centers and $n^2$ many distances, giving rise to at most $n^3$ many ranges.

We note that unlike for $k$-median, these bounds are not known to be tight for specific metrics. For example, finite $n$-point metrics admit coresets of size $\tilde{O}(k \cdot \log n \cdot \varepsilon^{-2})$ for $k$-means, which is optimal [18, 21], whereas the VC-dimension derived bound only yields a coreset of size $\tilde{O}(k \cdot \log n \cdot \varepsilon^{-3})$. Nevertheless, for the aforementioned clustering objectives such as shortest path metrics in planar graphs and the Frechet distance, this still yields improvements over the state of the art. In an earlier version of this paper, we claimed an improved bound of $\tilde{O}\left(k \cdot d_{\text{VC}} \cdot (\varepsilon^{-2} + \varepsilon^{-z})\right)$, which turned out to be false. We believe that the bounds in Theorem 1.2 are optimal in the sense that there exist metrics and ranges of $\varepsilon$ and $k$ for which the stated bound cannot be improved by any coreset construction. A discussion is given at the end of this paper.

## 1.2 Our Techniques and Related Work

**Coresets via Uniform Sampling.** We give explicit bounds of various papers studying coresets with bounded VC dimension in Table 1.

| Reference | Size (Number of Points) |
|---:|:---:|
| Feldman, Langberg (STOC'11) [28] | $\tilde{O}(k^3 \cdot d_{\text{VC},w} \cdot \varepsilon^{-2})$ |
| Munteanu, Schwiegelshohn, Sohler, and Woodruff (NeurIPS'18) [44] | $O(k^2 \cdot d_{\text{VC}} \cdot \varepsilon^{-2} \log n)$ |
| Feldman, Schmidt, Sohler (Sicomp'20) [30] | $\tilde{O}(k^2 \cdot d_{\text{VC},w} \cdot \varepsilon^{-2})$ |
| Braverman, Cohen-Addad, Jiang, Krauthgamer, Schwiegelshohn, Toftrup, Wu (FOCS'22) [11] | $\tilde{O}(k^3 \cdot d_{\text{VC}} \cdot \varepsilon^{-3})$ |
| Cohen-Addad, Saulpic, Schwiegelshohn (FOCS'23) [22] | $\tilde{O}(k^2 \cdot d_{\text{VC}} \cdot \varepsilon^{-3})$ |
| **This paper** | $O(k \cdot d_{\text{VC}} \cdot \varepsilon^{-2})$ |

Table 1: Comparison of VC dimension derived coreset sizes for $k$ median. $d_{\text{VC}}$ denotes the unweighted VC dimension and $d_{\text{VC},w}$ denotes the weighted VC dimension where the upper bound on the VC dimension has for the ranges $R(S,r) := \{p | w_p \cdot \text{dist}(p,\mathcal{S}) \geq r\}$ have to hold for any fixed choice of weights. Note that $d_{\text{VC},w} \geq d_{\text{VC}}$. polylog factors are omitted from all references. The result by [44] is also implicitly given by Chen [15].

The starting point of our work is Chen's seminal paper [15]. Most coreset algorithms for $k$-clustering start by computing a suitable constant factor approximation $\mathcal{A}$ to the underlying clustering objective. Chen adds the additional key preprocessing step that divides the input into groups $G$ with the property:

**Property 0.** Any two points have the same cost up to a constant factor.

Subsequently, Chen uses uniform sampling on every group. With this algorithm, there exists a simple reduction to approximating ranges induced by metric balls. For a solution $\mathcal{S}$, we use $R_G(\mathcal{S}, r) := \{p \in G \mid \text{dist}(p,\mathcal{S}) \geq r\}$, with $\mathcal{R}^k = \cup_{\mathcal{S},r} R(\mathcal{S}, r)$ denoting the range space induced by the $k$-fold intersection of metric balls. The cost with respect to any solution $\mathcal{S}$ can then be expressed as

$$\text{cost}(G,\mathcal{S}) = \sum_{p\in G} \text{cost}(p,\mathcal{S}) = \sum_{p\in G} \int_{r=0}^{\infty} \mathbb{1}_{\text{cost}(p,\mathcal{S})\geq r} dr = \int_{r=0}^{\infty} |R(\mathcal{S},r)| dr.$$

By the same token, the cost of a coreset $\Omega$ obtained from uniform sampling with weights set to $\frac{|G|}{|\Omega|}$ reduces to the expression

$$\text{cost}(\Omega,\mathcal{S}) = \sum_{p\in\Omega} \frac{|G|}{|\Omega|} \text{cost}(p,\mathcal{S}) = \int_{r=0}^{\infty} \frac{|G|}{|\Omega|} \cdot |R_\Omega(\mathcal{S},r)| dr.$$

With a clever charging argument, found for example in [30] but also implicit in Chen's work, we can then show that if, for all ranges $R$,

$$\frac{|G|}{|\Omega|} \cdot |R_\Omega(\mathcal{S},r)| = |R_G(\mathcal{S},r)| \pm \varepsilon \cdot |G|, \tag{1}$$

then $\text{cost}(\Omega,\mathcal{S}) \in (1\pm\varepsilon)\cdot \text{cost}(G,\mathcal{S})$ for all solutions $\mathcal{S}$.

Condition (1) is equivalent to an $\varepsilon$-approximation of a range space and is among the most well-studied objects in empirical processes. Indeed, the seminal paper by Vapnik and Chervonenkis [49] proved that taking $O\left(\frac{D}{\varepsilon^2}\log\left(D\varepsilon^{-1}\right)\right)$ independent uniform samples (with $D$ being the VC dimension of $\mathcal{R}$) leads to Condition (1) holding for all $R \in \mathcal{R}^k$. Subsequent improvements and refinements have been discovered, see Chazelle's book [14] for an overview. For $k$-clustering in particular, we have $D = O(kd\log k)$ if the VC dimension of $\cup_{c,r} R(c.r)$ is $d$, see [27]. This suggests that the coreset size is at most $\tilde{O}(kd\varepsilon^{-2})$. Unfortunately, Chen's reduction requires performing this procedure on $k\log n$ many groups, leading to an overall coreset of size $\tilde{O}(k^2 d\varepsilon^{-2}\log n)$. A fairly involved modification of Chen's reduction due to [11] replaced the dependency on $k\log n$ with a dependency poly$(k, \varepsilon^{-1})$.

**Coresets via Non-Uniform Sampling.** If we are given a good characterization of the number of clusterings, significantly better results are achievable via non-uniform sampling. Crucially, non-unform sampling requires no preprocessing other than computing the constant factor approximation $\mathcal{A}$ and thus has no loss in parameters. Unfortunately, the aforementioned reduction that allows us to gather points by way of $|R(\mathcal{S}, r)|$ no longer works as non-uniform sampling necessitates non-uniform weighting. Previous work now faced the option of dealing with weighted analogues of the range spaces. These are often not as well understood as their unweighted counterparts (for example, shortest path metrics in planar graphs [8]). Instead, most recent work relied on ad-hoc enumerations of all candidate solutions in the underlying metric, see [4, 18, 21, 39].

**Group Sampling.**

At this stage, we consider the Group Sampling algorithm by [18]. It relaxes the preprocessing of Chen's algorithm by only requiring that for every group $G$

**Property 1.** Any two clusters in $G$ have the same cost up to a constant factor.

**Property 2.** Any two points from the same cluster have the same cost up to a constant factor.

Notice that properties 1 and 2 relax Chen's property 0 by requiring that they only apply for points in the same cluster, while enforcing a very similar property to that of Chen onto the clusters as a whole. Crucially, this requirement drastically reduces the number of groups from $k\log n$ to polylog$(\varepsilon^{-1})$. The sampling procedure then first picks a cluster uniformly at random and then samples a points from the picked cluster also uniformly at random. Unfortunately, it is not known how to use the VC dimension in combination with this algorithm, without accounting for the number of distinct weights. Since there can be as many as $k$ different weights, it is not clear how to improve over a $\tilde{O}(k^2 d\varepsilon^{-2})$ coreset bound.

Our main contribution is to augment the Group Sampling algorithm by another property:

**Property 3.** Any two clusters in $G$ have that the ratio of their average costs is either between 1/2 and 2, or very small, or very large.

We call this the *layering* property. To illustrate why this property is useful, we consider the simplified case of having only average costs $V^i$ for $i \in \{0, \ldots\ldots, k-1\}$ in a group $G$, with $V$ being a very large number, possibly of the order poly$(k, \varepsilon^{-1})$. Denote by $G_i$ the subgroup clusters with average cost $V^i$. When restricted to the subgroup $G_i$, the Group Sampling algorithm is merely uniform sampling. In this case we can use the VC dimension to readily enumerate over all candidate clusterings with $k$ centers. Specifically, enumerating the number of such clusterings to a desired precision yields $|G_i|^{O(kd)}$ different candidate solution. If we were to combine these candidate clusterings naively via $\prod_{i=0}^{k-1} |G_{i-1}|^{O(kd)}$, we would obtain $n^{O(k^2 d)}$ clusterings in total, leaving us with a $\tilde{O}(k^2 d\varepsilon^{-2})$ coreset bound.

The main observation is that a significantly tighter combination is possible. Specifically, we enumerate over the number of solutions incrementally, starting from the clusters in subgroup $G_0$ with smallest average cost. Since the points in subgroup $G_0$ have a very small cost compared to any subgroup $G_i$, $i > 0$, any center used to serve points in $G_0$ cannot induce many different ranges of cost for points in $G_i$ for $i > 0$. Conversely, if centers used to induce many different costs in $G_i$, most of them cannot serve points in $G_0$. In a nutshell, the layering property either forces no interaction between subgroups to take place, reducing the total number of centers for each subgroup, or the ranges for a subgroup $G_i$ are determined by the ranges in subgroups $\cup_{j\leq i-1} G_j$. This allows us to characterize the clusters in an optimal way, yielding a nearly tight bound of $n^{\tilde{O}(kd)}$ many clusterings for any group with the layering property. The layering property itself is inexpensive to enforce, barely increasing the number of groups. We believe that it is a useful property on its own and might have applications beyond the aims of this paper.

### 1.3 Further Related Work

Coresets for clustering have a long and rich history and we briefly highlight some of the very important works not covered above. The earliest coreset examples relied heavily on techniques from computational geometry [32, 33], which, despite falling out of fashion, are still relevant today for designing coresets for constrained clustering objectives [9, 37, 46] or in low-dimensional settings [1, 38]. Chen's seminal work [15] introduced sampling techniques to coreset construction and drew the connection to learning theory, a connection that was further explored in [28, 40]. Subsequently, most coreset research explored Euclidean spaces by way of dimension reduction [6, 19, 30, 35, 47].

Aside from Euclidean spaces, coreset research has also studied doubling metrics [36], the Frechet metric for clustering time series [24], shortest path distances in minor free graphs [10] and geometric intersection graphs [4], and bounded treewidth graphs [2]. Further results also focus on obtaining fast running times [25], streaming [23], distributed computing [3], or dynamic algorithms [34].

## 2 Framework

As is usual with coreset algorithms, we start by computing a constant factor approximation $\mathcal{A}$. The overall coreset error will be $\varepsilon \cdot (\text{cost}(P, \mathcal{A}) + \text{cost}(P, \mathcal{S}))$ for any solution $\mathcal{S}$. This is $\varepsilon \cdot \text{cost}(P, \mathcal{S})$ after rescaling $\varepsilon$ by constant factors, as we assume $\mathcal{A}$ to be a constant-factor approximation of the optimal solution.

### 2.1 Preliminaries

Let $P$ be a set of points in a metric space $(X, \text{dist})$ and a constant factor approximation $\mathcal{A}$ of the $(k, z)$-clustering objective for point set $P$ on space $(X, \text{dist})$. [3] The cost of a point $p$ in solution $\mathcal{S}$ is $\text{cost}(p, \mathcal{S}) := \min_{c \in \mathcal{S}} \text{dist}^z(p, c)$. We refer to a cluster in the approximation as $C_i$, which represents the set of points that are closest to center $c_i \in \mathcal{A}$.

Throughout the text, we use $\gamma$ to refer to some sufficiently large absolute constant. We furthermore use subscripts when it is necessary to differentiate between the $\gamma$ terms. Further, we will write $a \lesssim b$ if there exists a sufficiently large constant $\gamma$ such that $a \leq \gamma \cdot b$. Throught this paper, we will treat $z$ as a constant, that is $a \leq 2^{\text{poly}(z)} \cdot b$ will be written as $a \lesssim b$.

For readability's sake, we recall the definitions of shattering and combinatorial notion of dimension both in uniform and scale-sensitive settings. The definitions are taken from Chapter 7 in [48]. Let $X$ be a set of points and let $\mathcal{F}$ be a set of functions over $X$. We say that $\mathcal{F}$ shatters $Y \subseteq X$ if for every $Z \subseteq Y$ there exists a function $f \in \mathcal{F}$ s.t. $f(x) = \begin{cases} 1, & \text{if } x \in Z \\ 0, & \text{if } x \in Y \setminus Z \end{cases}$. We say that the VC dimension of $\mathcal{F}$ is the size of the largest shattered subset of $X$, and denote it by $d_{\text{VC}}^{\mathcal{F}}$.

### 2.2 Partitioning an Instance into Groups: Definitions

The algorithm partitions the input points into structured groups described as follows:

**Rings.** Let $\Delta_{C_i} = \frac{\text{cost}(C_i, \mathcal{A})}{|C_i|}$ be the average cluster cost and define for each cluster:

- Ring $R_{ij} = \{p \in C_i \mid 2^j \Delta_{C_i} \leq \text{cost}(p, \mathcal{A}) \leq 2^{j+1} \Delta_{C_i}\}$, i.e., the set of points that have roughly the same multiple (up to a factor of 2) of the average point's cost in that cluster in the approximate solution.
- Inner ring $R_I(C_i) = \bigcup_{j \leq \log(\varepsilon^z/\gamma)} R_{ij}$, i.e., the set of points whose cost in the approximate solution is significantly smaller than the average point's cost in that cluster. We write $R_I = \bigcup_{i \in [k]} R_I(C_i)$.
- Outer ring $R_O(C_i) = \bigcup_{j > \log(\gamma \cdot k^3/\varepsilon^{2z})} R_{ij}$, i.e., the set of points whose cost in the approximate solution is much larger than the average cost of the cluster they belong to. We write $R_O = \bigcup_{i \in [k]} R_O(C_i)$.

[3] The claims can be easily extended to approximation factors $\alpha$ with $\beta \cdot k$ many points. In order to minimize the parameters, we streamline the analysis.

- All those rings that are neither inner nor outer are the main rings. In particular, we let $R_j = \bigcup_{i \in [k]} R_{ij}$, that is the set of points that contribute roughly the same amount of cost to their respective cluster centers.

We may assume that the number of rings is at most $M \in \tilde{O}\left(k^3 \varepsilon^{-3} d_{\mathrm{VC}}\right)$, using the previous algorithm by [11]. The fundamental property of any ring $R_{ij}$ is that every point in it contributes roughly the same amount to $\mathrm{cost}(\mathcal{A})$.

**Layers.** Let $\phi$ be a power of 2 of the order $\left(\frac{k}{\varepsilon}\right)^{-10z} \cdot \gamma$, for a sufficiently large absolute constant $\gamma$ that depends on $z$ (but is independent of other problem parameters). Let $\ell \in \{0, 1, 2 \ldots \log \phi - 1\}$. We introduce a further refinement of rings that we call *layers*. We say that ring $R_{ij}$ belongs to *layer* $L_\ell$ if $2^\ell \cdot \phi^a \leq 2^j \cdot \Delta_{C_i} < 2^{\ell+1} \cdot \phi^a$ for some integer $a$.

**Groups.** Each group $G$ is comprised of sets of rings. Consider the set of $j$-th rings $R_j$ around the centers in $\mathcal{A}$. Then each ring $R_{ij}$ has some total cost over all of its points. A group $G_{jb}$ is then the subset of these $j$-th rings that have roughly the same total cost (where the $b$ subscript determines how large this cost is). This means that a group $G_{jb}$ is comprised of the $j$-th rings around some subset of the centers in our approximate solution. We first consider the main rings. For each $j$, the main rings $R_{i,j}$ are gathered into *main groups* $G_{j,b,\ell}$ defined as follows:

$$G_{j,b,\ell} := \left\{ p \mid \exists i,\ p \in R_{i,j} \in L_\ell \text{ and } \left(\frac{\varepsilon}{4z}\right)^z \cdot \frac{\mathrm{cost}(R_j, \mathcal{A})}{k} \cdot 2^b \leq \mathrm{cost}(R_{i,j}, \mathcal{A}) \leq \left(\frac{\varepsilon}{4z}\right)^z \cdot 2^{b+1} \cdot \frac{\mathrm{cost}(R_j, \mathcal{A})}{k} \right\},$$

where $b \in \{0, \ldots, \log(\gamma \cdot k/\varepsilon^z)\}$. For any $j$, let $G_{j,\min,\ell} := \cup_{b \leq 0, \ell} G_{j,b}$ be the union of the cheapest groups.

The set of outer rings is also partitioned into *outer groups*.

$$G^O_{b,\ell} = \Big\{ p \mid \exists i,\ p \in C_i \cap L_\ell \text{ and } \left(\frac{\varepsilon}{4z}\right)^z \cdot \frac{\mathrm{cost}(R^{\mathcal{A}}_O \cap L_\ell, \mathcal{A})}{k} \cdot 2^b \leq \mathrm{cost}(R_O(C_i) \cap L_\ell, \mathcal{A}) \leq \left(\frac{\varepsilon}{4z}\right)^z \cdot 2^{b+1} \cdot \frac{\mathrm{cost}(R^{\mathcal{A}}_O \cap L_\ell, \mathcal{A})}{k} \Big\},$$

where $b \in \{0, \ldots, \log(\gamma \cdot k/\varepsilon^z)\}$. We let as well $G^O_{\min} = \cup_{b \leq 0} G^O_{b,\ell}$.

With the definition of the groups, let us review the properties from Section 1.2. Properties 1 and 2 are still satisfied, as they are satisfied by the original group sampling algorithm and any subdivision of groups inherits these properties, so we focus on Property 3. Here, we intersect every group with the layers. By definition of a layer, the cost of a ring in $G_{j,b,\ell}$ satisfied $2^\ell \cdot \phi^a \leq 2^j \cdot \Delta_i \leq 2^{\ell+1} \cdot \phi^a$. Therefore, the ratio of $\frac{2^j \cdot \Delta_i}{2^{j'} \cdot \Delta_{i'}}$, for two rings $R_{ij}$ and $R_{i'j'}$, is either within a factor of $[1/2, 2]$, or the ratio is either at least $\phi$ or at most $\phi^{-1}$. The cost of the points in their corresponding ring is within a factor 2 of the value of $2^j \cdot \Delta_i$, so the ratio of the costs of points is within a factor $[1/4, 4]$, or at least $\phi/2$ or at most $2\phi^{-1}$.

## 2.3 The Layered Group Sampling Algorithm

Following the computation of $\mathcal{A}$, for which one could use any suitable approximation algorithm, we perform the portioning into groups. In particular, we have the following:

**Proposition 2.1.** *The total number of layered groups is in $O(\log^3(k\varepsilon^{-1}))$.*

*Proof.* There are $O(\log(k\varepsilon^{-1}))$ many different values of $j$ for which we consider rings $R_{i,j}$ and there are $O(\log(k\varepsilon^{-1})$ values of $b$. Further, there are $O(\log \phi) = O(\log(k\varepsilon^{-1}))$ many values of $\ell$. Each group is identified by a combination of these values, so we have $O(\log^3(k\varepsilon^{-1}))$ many groups. □

For each group $G_{j,b,\ell}$, we sample a point $p \in C_i$ with probability

$$\mathbb{P}[p] = \frac{\mathrm{cost}(C_i \cap G_{j,b,\ell}, \mathcal{A})}{\mathrm{cost}(G_{j,b,\ell}, \mathcal{A})} \cdot \frac{1}{|C_i \cap G_{j,b,\ell}|}.$$

**Algorithm 1** Layered Group Sampling

Compute a constant factor approximation $\mathcal{A}$
Partition the points into main, outer, and inner groups.
**for** Inner Groups $G$ **do**
  **for** $C_i$ **do**
    Add $c_i \in \mathcal{A}$ with weight $G \cap C_i$ to $\Omega$.
**for** Main Groups $G$ **do**
  For every $p \in C_i \cap G$, set $\mathbb{P}[p] \propto \frac{1}{k} \cdot \frac{1}{|C_i \cap G|}$ and $w_p = \frac{1}{\mathbb{P}[p]}$
  Add $m$ points independently sampled with replacement and weighted by $\frac{w_p}{m}$ to $\Omega$.
**for** Outer Groups $G$ **do**
  For every $p \in C_i \cap G$, set $\mathbb{P}[p] \propto \text{cost}(p, \mathcal{A})$ and $w_p = \frac{1}{\mathbb{P}[p]}$
  Add $m$ points independently sampled with replacement and weighted by $\frac{w_p}{m}$ to $\Omega$.
Return $\Omega$

For each group $G^O_{j,\ell}$, we sample a point $p \in C_i$ with probability

$$\mathbb{P}[p] = \frac{\text{cost}(p, \mathcal{A})}{\text{cost}(G_{j,b,\ell}, \mathcal{A})}.$$

The only modification to the Group Sampling algorithm by [18] is the addition of layers in the preprocessing. The sampling distribution stays the same.

## 2.4 Outline of the Analysis

We will follow the outline of a chaining argument used in [20]. The main steps of the arguments are (i) reduction to a Gaussian process, (ii) bounding the variance of the Gaussian process, and (iii) bounds on net sizes. The nets, which are our main contribution, are given in Section 3. The variances and the reduction to Gaussian processes, whenever appropriate, are taken from [20] and repeated in Section 4 for completeness.

We will work with the designated coreset over groups $G$:

$$\sum_G \frac{1}{|\Omega_G|} \sum_{p \in \Omega_G} w_p \text{cost}(p, \mathcal{S}).$$

where group $G$ corresponds to one of the groups defined above, $\Omega_G$ is the sketch we construct for $G$ and $w_p$ is the weight associated with $p$ in $G$.

The coreset guarantee we ultimately aim to prove is:

$$|\text{cost}(\Omega_G, \mathcal{S}) - \text{cost}(G, \mathcal{S})| \le \frac{\varepsilon}{\log^3(k\varepsilon^{-1})} \cdot (\text{cost}(P, \mathcal{S}) + \text{cost}(P, \mathcal{A})). \tag{2}$$

Due to Proposition 2.1, this implies an $O(\varepsilon)$ coreset in general.[4] Equation (2) is equivalent to showing

$$\sup_{\mathcal{S}} \frac{|\text{cost}(\Omega, \mathcal{S}) - \|v^{\mathcal{S}}\|_1|}{\text{cost}(P, \mathcal{S}) + \text{cost}(P, \mathcal{A})} \le \frac{\varepsilon}{\log^3(k\varepsilon^{-1})},$$

where $v^{\mathcal{S}} \in \mathbb{R}^n$ is the vector of per-point costs with respect to the candidate solution $\mathcal{S}$. That is, $v_i = \text{cost}(p_i, \mathcal{S})$. In fact, our goal is to show that

$$\mathbb{E}_\Omega \left[ \sup_{\mathcal{S}} \frac{|\text{cost}(\Omega, \mathcal{S}) - \|v^{\mathcal{S}}\|_1|}{\text{cost}(P, \mathcal{S}) + \text{cost}(P, \mathcal{A})} \right] \le \frac{\varepsilon}{\log^3(k\varepsilon^{-1})}, \tag{3}$$

where $\mathbb{E}_\Omega$ is meant to denote the expectation over the randomness of $\Omega$. This implies that the desired guarantee holds with constant probability.

We partition the clusters of main group in any solution $\mathcal{S}$ by type.

[4] The dependency on the log factors can be improved, for example a tighter bound of $\log \varepsilon^{-1} \cdot \log^2(k\varepsilon^{-1})$ is also possible with less effort, and the $\log k$ dependencies might be avoidable entirely. We did not attempt to optimize these dependencies.

**Definition 2.2** (Clusters of Type $i$)**.** *Given approximate solution $\mathcal{A}$ and candidate solution $\mathcal{S}$, let $c$ be any center in $\mathcal{A}$. For $C = \{p \in P \mid c = \arg\min_{c' \in C} \operatorname{dist}(p, c')\}$, we define $q_p^{\mathcal{S}} = \min_{p' \in C} \operatorname{cost}(p', \mathcal{S})$, the minimum cost to solution $\mathcal{S}$ in $p$'s cluster, simply writing $q_p$ if $\mathcal{S}$ is clear from context. Furthermore, we define type $i$ to be the set of clusters $C$ such that $q_p \in [2^i, 2^{i+1}) \cdot \min_{p' \in C} \operatorname{cost}(p', \mathcal{A})$.*

In words, a cluster's type with respect to a candidate solution $\mathcal{S}$ measures how expensive it is with respect to the candidate solution $\mathcal{S}$ compared to its cost in the constant-factor approximation $\mathcal{A}$. The number of clusters $C_j \in T_i$ is denoted by $k_i$. If $C_j$ is of type $i \le 2z$, we say $C_j$ is of type $T_{small}$ and if $C_j$ is of type $i \ge \log \gamma\varepsilon^{-z}$, for a sufficiently large absolute constant $\gamma$, we say that $C_j$ is of type $T_{large}$. Similarly, we also partition the clusters of an outer group in any solution $\mathcal{S}$ by type. We consider a cluster $C_j$ to be in $T_{close}$ if $C_j$ is of type $i \le 2z$, otherwise we say $C_j$ is of type $T_{far}$. In the following, let $v_{small}^{\mathcal{S}}$, $v_i^{\mathcal{S}}$, $v_{close}^{\mathcal{S}}$ and $v_{far}^{\mathcal{S}}$ be the projection of $v^{\mathcal{S}}$ onto the coordinates induced by $T_{small}$, $T_i$, $T_{close}$ and $T_{far}$, respectively. We also use $v_{j,large}^{\mathcal{S}}$ to denote the projection of $v^{\mathcal{S}}$ onto the coordinates of points in a cluster $C_j \in T_{large}$.

Then, we show

$$\mathbb{E}_\Omega\left[\sup_{\mathcal{S}}\left|\frac{\frac{1}{|\Omega|}\sum_{C_j \in T_{small}}\sum_{p \in C_j \cap \Omega} w_p \operatorname{cost}(p,\mathcal{S}) - \|v_{small}^{\mathcal{S}}\|_1}{\operatorname{cost}(P,\mathcal{S}) + \operatorname{cost}(P,\mathcal{A})}\right|\right] \le \frac{\varepsilon}{\log^4(k\varepsilon^{-1})} \tag{4}$$

$$\mathbb{E}_\Omega\left[\sup_{\mathcal{S}}\left|\frac{\frac{1}{|\Omega|}\sum_{C_j \in T_i}\sum_{p \in C_j \cap \Omega} w_p \operatorname{cost}(p,\mathcal{S}) - \|v_i^{\mathcal{S}}\|_1}{\operatorname{cost}(P,\mathcal{S}) + \operatorname{cost}(P,\mathcal{A})}\right|\right] \le \frac{\varepsilon}{\log^4(k\varepsilon^{-1})} \tag{5}$$

$$\mathbb{E}_\Omega\left[\sup_{\mathcal{S}}\left|\frac{\frac{1}{|\Omega|}\sum_{p \in (C_j \in T_{large}) \cap \Omega} w_p \operatorname{cost}(p,\mathcal{S})}{\operatorname{cost}(\cup_{C_j \in T} C_j, \mathcal{S})} - 1\right|\right] \le \varepsilon \tag{6}$$

$$\mathbb{E}_\Omega\left[\sup_{\mathcal{S}}\left|\frac{\frac{1}{|\Omega|}\sum_{C_j \in T_{close}}\sum_{p \in C_j \cap \Omega} w_p \operatorname{cost}(p,\mathcal{S}) - \|v_{close}^{\mathcal{S}}\|_1}{\operatorname{cost}(P,\mathcal{S}) + \operatorname{cost}(P,\mathcal{A})}\right|\right] \le \frac{\varepsilon}{\log^4(k\varepsilon^{-1})} \tag{7}$$

$$\sup_{\mathcal{S}}\left|\frac{\frac{1}{|\Omega|}\sum_{C_j \in T_{far}}\sum_{p \in C_j \cap \Omega} w_p \operatorname{cost}(p,\mathcal{S}) - \|v_{far}^{\mathcal{S}}\|_1}{\operatorname{cost}(P,\mathcal{S}) + \operatorname{cost}(P,\mathcal{A})}\right| \le \frac{\varepsilon}{\log^4(k\varepsilon^{-1})} \tag{8}$$

Note that if Equation (5) holds for $i \in \{2z, \ldots, \log(\gamma\varepsilon^{-z})\}$, then Equation (3) holds, as the error from each type can only sum up in the worst case and there are at most $O(\log \varepsilon^{-1})$ many types.

We can now show that we obtain a coreset by way of the following lemmas. Dependencies $f(z)$, for functions that are independent of problem parameters other than $z$ are omitted for readability.

**Lemma 2.3.** *Fix a main group $G$. Let $|\Omega| \in \tilde{\Theta}(kd_{\mathrm{VC}}\varepsilon^{-2})$. Then Equation (4) holds.*

**Lemma 2.4.** *Fix a main group $G$. Let $|\Omega| \in \tilde{\Theta}(kd_{\mathrm{VC}}\varepsilon^{-2}\min(\varepsilon^{-z+1}, k))$. Then Equation (5) holds.*

**Lemma 2.5.** *Fix a main group $G$. Let $|\Omega| \in \tilde{\Theta}(k\varepsilon^{-2})$. Then Equation (6) holds.*

**Lemma 2.6.** *Fix an outer group $G$. Let $|\Omega| \in \tilde{\Theta}(kd_{\mathrm{VC}}\varepsilon^{-2})$. Then Equation (7) holds.*

**Lemma 2.7.** *Fix an outer group $G$. Then Equation (8) holds.*

**Remark 2.8.** *The bound for Lemma 2.5 could also be proved as $\mathbb{E}_\Omega\left[\sup_{\mathcal{S}}\left|\frac{\frac{1}{|\Omega|}\sum_{C_j \in T_{large}}\sum_{p \in C_j \cap \Omega} w_p \operatorname{cost}(p,\mathcal{S}) - \|v_{large}^{\mathcal{S}}\|_1}{\operatorname{cost}(P,\mathcal{S}) + \operatorname{cost}(P,\mathcal{A})}\right|\right] \le \varepsilon$. However, this yields a $\log^z(k)$ dependency. While we have $\log^{O(1)}(k)$ dependencies that we did not try to optimize, we wished to avoid a $\log^z(k)$. The proof difficulty does not change substantially. In addition, we will show that the bound for Lemma 2.7 holds deterministically by construction of the groups.*

Finally, the following lemma states that the cheap groups $G_{j,\min}$ and $G_{\min}^O$ have simple coresets: one simply weighs each center in approximate solution $\mathcal{A}$ by the cardinality of the cheap groups.

**Lemma 2.9** (Lemma 4 in [18])**.** *Let $G_{cheap} = G^O_{min} \cup \left(\bigcup_j G_{j,min}\right)$ be the union of the cheapest main and outer groups, as defined in Section 2.2. For each cluster $C$ with center $c$ in $\mathcal{A}$, let $w(c) = |G_{cheap} \cap C|$. Then $\left|\sum_{p\in G_{cheap}} \text{cost}(p,\mathcal{S}) - \sum_{c\in\mathcal{A}} w(c)\cdot\text{cost}(c,\mathcal{S})\right| \leq \varepsilon\cdot\text{cost}(P,\mathcal{S})$.*

Using these lemmas, we can prove our main theorem:

*Proof of Theorem 1.2.* The total number of types is of the order $O(\log(\varepsilon^{-1}))$ and by Proposition 2.1, the total number of groups is of the order $O(\log^3(k\varepsilon^{-1}))$. The expected total error when adding up the error bounds from Lemmas 2.3, 2.4, 2.5, 2.6, 2.7 and 2.9 is then $\frac{\varepsilon}{\log^4(k\varepsilon^{-1})}\cdot O(\log^4(k\varepsilon^{-1}))\,(\text{cost}(P,\mathcal{S}) + \text{cost}(P,\mathcal{A})) = O(\varepsilon)\cdot\text{cost}(P,\mathcal{S})$ for all solutions $\mathcal{S}$, as we assumed $\mathcal{A}$ is a constant factor approximation. To obtain a constant probability of success, use Markov's inequality. The space bound likewise follows by combining the coresets for all groups, each of which has size at most $\tilde{O}(kd_{\text{VC}}\varepsilon^{-2}\cdot\min(\varepsilon^{-z+1},k)))$. □

# 3 Clustering Nets

The purpose of this section is to introduce the tools with which we will discretize the space of costs of the candidate solutions. Traditionally, coreset literature has done this using approximate centroid sets [29, 18]. These are a direct discretization of the set of candidate solutions up to some permitted error $\alpha$. In this paper, we use an equivalent notion of *clustering nets*, as first introduced by Cohen-Addad et al. [21]. Specifically, let $v \in \mathbb{R}^n$ be any vector such that $v_p = \text{cost}(p,\mathcal{S})$ for some candidate solution $\mathcal{S}$. Consequently, the space of all possible $v$ vectors represents all of the costs which candidate solutions may incur.

Our goal will be to provide a small discretization $\mathcal{N}$ of these such that, for any vector $v$, there exists an element $v' \in \mathcal{N}$ such that $|v - v'| \leq \alpha\Lambda$, where $\alpha$ is a precision parameter and $\Lambda$ is the range with respect to which we are preserving the error. In essence, one can think of $\Lambda$ as being on the order of $\text{cost}(p,\mathcal{S})$. Specifically, Lemma 3.4 shows that we can bound the errors with respect to $\text{cost}(p,\mathcal{A})$ on clusters in type $i$ by bounding errors with respect to $\text{dist}^{z-1}(p,\mathcal{S})\cdot\text{dist}(p,\mathcal{A})$.

**Definition 3.1** ($(\alpha, k, 2^i)$-clustering net)**.** *Let $(X,\text{dist})$ be a metric space, $P$ a set of points, $k$ a positive integer and let $\alpha \in (0,\frac{1}{2})$ be a precision parameter. For a given (optimal or approximate) solution $\mathcal{A}$, let $G$ be a group. Let $\mathcal{C} \in (X,\text{dist})^k$ be a set of (potentially infinite) $k$-clusterings for the $(k,z)$-clustering objective. We say that a set of cost vectors $\mathcal{N}_P(\alpha,k,2^i) \in \mathbb{R}^{|P|}$ is a type $i$ clustering net if, for every solution $\mathcal{S}\in\mathcal{C}$, there exists a vector $v\in\mathcal{N}_P(\alpha,k,2^i)$ such that the following conditions hold:*

$$|v_p - \text{cost}(p,\mathcal{S})| \leq \alpha\cdot(\text{dist}^{z-1}(p,\mathcal{S})\cdot\text{dist}(p,\mathcal{A}) + \text{dist}^z(p,\mathcal{A})), \qquad \forall p \in \textit{type } i \textit{ cluster}$$
$$v_p = 0, \qquad \forall p \notin \textit{type } i \textit{ cluster.}$$

In essence, this is a net over the solution space for clusters of type $i$. For the types $T_{small}$ or $T_{close}$, the error guarantee reduces to $\alpha\cdot\text{dist}^z(p,\mathcal{A})$.

We will frequently use the following three lemmas.

**Lemma 3.2** (Triangle Inequality for Powers [42])**.** *Let $a,b,c$ be an arbitrary set of points in a metric space with distance function $d$ and let $z$ be a positive integer. Then for any $\beta > 0$,*

$$\text{dist}^z(a,b) \leq (1+\beta)^{z-1}\text{dist}^z(a,c) + \left(1+\frac{1}{\beta}\right)^{z-1}\text{dist}^z(b,c)$$
$$|\text{dist}^z(a,\mathcal{S}) - \text{dist}^z(b,\mathcal{S})| \leq \beta\text{dist}^z(a,\mathcal{S}) + \left(1+\frac{2z}{\beta}\right)^{z-1}\text{dist}^z(a,b).$$

We now apply this to prove the following two lemmas:

**Lemma 3.3.** *Let $\mathcal{A}$ be an approximate solution, $C$ be any cluster from this solution, and $G$ be a layered group obtained from $\mathcal{A}$. Let $p$ be any point in cluster $C\cap G$. Furthermore, assume $C$ is in type $i$ with respect to candidate solution $\mathcal{S}$. Then for some sufficiently large constant $\gamma_1$, we have*

$$\text{cost}(p,\mathcal{S}) \leq \gamma_1\cdot(2^{i/z}\text{dist}^{z-1}(p\mathcal{S})\text{dist}(p,\mathcal{A}) + \text{dist}^z(p,\mathcal{A})).$$

**Lemma 3.4.** *Let $\mathcal{A}$ be an approximate solution, $C$ be any cluster from this solution, and $G$ be a layered group obtained from $\mathcal{A}$. Let $p$ be any point in cluster $C \cap G$. Furthermore, assume $C$ is in type $i$ with respect to candidate solution $\mathcal{S}$. Assume also that $2^z < \gamma_1$ for some constant $\gamma_1$. Then*

$$\text{cost}(p, \mathcal{S}) - q_p \leq (2 + \gamma_1) \cdot 2^{i(1-1/z)} \cdot \text{cost}(p, \mathcal{A}) \leq (2 + \gamma_1) \cdot \text{dist}(p, \mathcal{S})^{z-1} \cdot \text{dist}(p, \mathcal{A}).$$

The proofs of these lemmas can be found in Appendix A.1 and A.2, respectively. In what follows, we introduce the key notion by which we characterize the solutions. This definition is a straightforward extension of approximate centroid sets from [18], albeit with $k$ centers instead of just one.

Within a group, we sort the clusters by increasing $\Delta_i$, that is $\Delta_1 = \min_g \Delta_{C_g}$ and $\Delta_m = \max_g \Delta_{C_g}$. We show that it is possible to combine their clustering nets for each $\Delta_g$ to obtain a clustering net for the group in its entirety.

We define interaction profiles for solutions $\mathcal{S}$ as a $k$ tuple $I(\mathcal{S}) \in \{0, 1, \ldots, k\}^m$ with $\sum_{i=1}^{k} I(\mathcal{S})_g \leq k$, where $m = k$ for main groups and $m = |\Omega_G|$ for outer groups. We use $I(\mathcal{S})_1$ to denote the number of centers in $\mathcal{S}$ with $\text{cost}(p, s) \leq r \cdot \text{cost}(p, \mathcal{A})$ for some point $p \in A_1$, where $s \in \mathcal{S}$. Recursively, define $I(\mathcal{S})_g$ to be the number of centers in $\mathcal{S}$ s.t. $\text{cost}(p, s) \leq \gamma \cdot 2^i \cdot \text{cost}(p, \mathcal{A})$, where $s \in \mathcal{S}$, for some point $p \in U_g$ and a sufficiently large constant $\gamma$ that depends on $z$, but is indepenedent of the other parameters. The number of interaction profiles are bounded as follows.

**Observation 3.5** (Number of Interaction Profiles)**.** *There are at most $(2e)^k$ interaction profiles for a main group $G$, and at most $(2e|\Omega_G|)^k$ interaction profiles for an outer group $G$ with sample $\Omega_G$.*

*Proof.* By the "stars and bars" principle (see e.g., Flajolet and Sedgewick [31, Chapter I.3]), we have that the number of possible ways $h$ nonnegative integers sum to some at most another integer $H$ is $\binom{H+h}{h}$.

In the case of a main group $G$, as $I_i$ is the number of centers serving the $i$-th cluster in sorted order, and $\sum_{i \in [k]} I_i \leq k$, we have that the number of interaction profiles for main groups is $\binom{2k}{k} \leq (2e)^k$.

For outer groups, we first select $k$ rings where we place the centers. Since there are at most $|\Omega_G|$ many rings, there are $\binom{|\Omega_G|}{k} \leq |\Omega_G|^k$ many such selections. For each selection, the stars and bars principle applies, which yields a total number of $(2e|\Omega_G|)^k$ many interaction profiles. ☐

## 3.1 Clustering Nets for Main Groups

We consider the clusters composing a main group $G_{j,b,\ell}$, or simply $G$ for the purpose of this section. Let us recall that their average costs are related as follows due to layering property: Either $\Delta_{g+1} \geq \phi/2 \cdot \Delta_g$ or $\Delta_{g+1} \leq 4 \cdot \Delta_g$.

We prove the following lemma:

**Lemma 3.6.** *For any main group $G$ and a sample $\Omega_G$ and any $\alpha \geq \varepsilon^{2z+2}/k$, there exists a type $i$ clustering net $\mathcal{N}_G(\alpha, k, 2^i)$ for $G$ of size*

$$\left|\mathcal{N}_G(\alpha, k, 2^i)\right| \leq \exp\left(\gamma \cdot \alpha^{-1} \cdot (2^{i/z} + 1) \cdot k \log k \cdot d_{\text{VC}} \cdot \log |\Omega_G|\right),$$

*for some absolute constant $\gamma > 0$.*

We proceed with the following steps. First, we assume without loss of generality that the clusters are sorted by increasing value $\Delta_{C_g}$, that is $C_1$ has the smallest average cost and $C_k$ has the maximum average cost. Further, we let $A_1$ be the set composed of the union of all clusters with $\Delta_g \leq 4\Delta_1$. We recursively define $U_j = \cup_{j'=1}^{j-1} A_{j'}$, with $U_1 = \emptyset$ and $A_j = G \setminus U_j$, that is $A_2$ is the set composed of the clusters $C_{g'}$ with $\Delta_{g'} \leq 4 \cdot \min_{C_g \in G \setminus A_1} \Delta_g$.

The clustering net construction idea consists of repeatedly applying the following procedure for a single group $G$: Every cluster union composing the group is only allowed to be served by centroids in its own clustering net or by the ones of cluster unions preceding it, as depicted in Figure 1.

In Claim 3.7, we prove that there exists a clustering net of small enough size for clusters with the same average cost (up to a constant) when they are considered in isolation. This is extended to all centers in Claim 3.8.

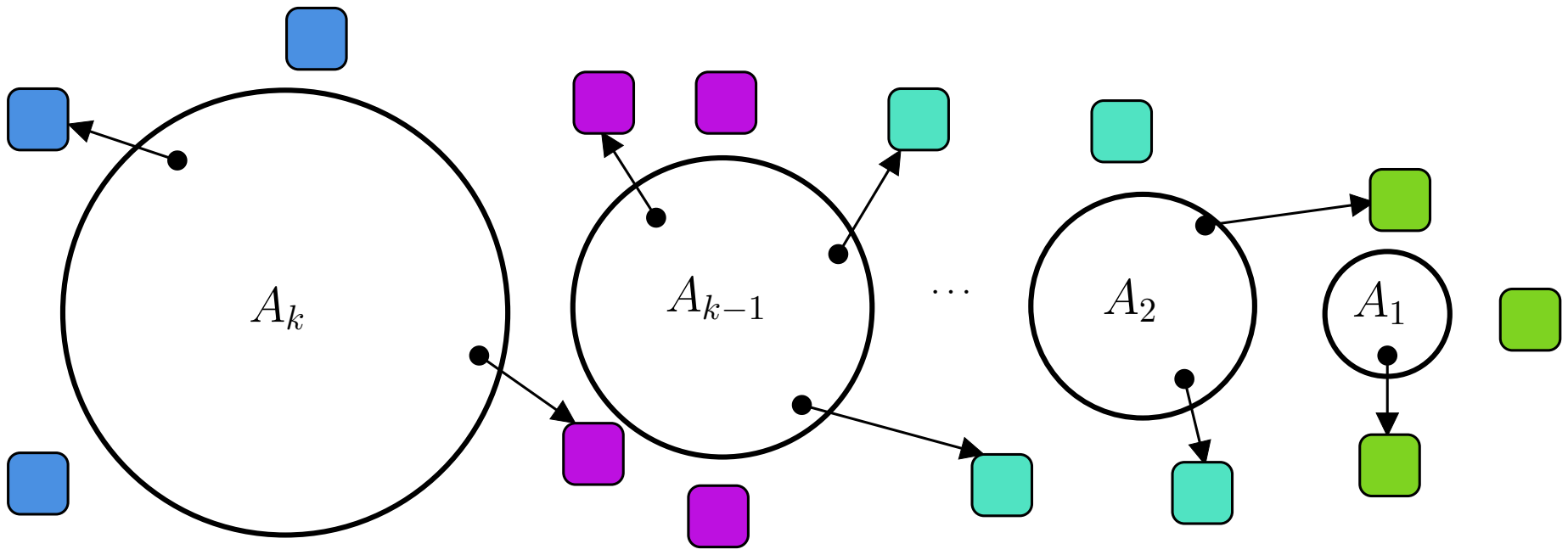

Figure 1: We draw one layered group $G$, composed of points in various $A_g$'s, i.e., for each $i$ points may belong to different clusters all with roughly the same average cost. Points in $A_g$ is only allowed to be served by centroids in the approximate centroid sets of $A_j$'s for $j \leq g$ (color-coded squares in the figure).

**Clustering Net for a Single Cluster Union $A_j$** In this part, we only focus on a single cluster union $A_j$ for group $G$, composed of clusters with equivalent (up to a constant) average cost. We prove the following claim:

**Claim 3.7.** *Fix a group $G$ with a sample $\Omega_G$, a cluster union $A_j$, and an $I_j$ being the $j$th entry in an interaction profile $I$. Then, the size of a type $i$ clustering net $\mathcal{N}_{A_j}(\alpha, I_j, 2^i)$ is bounded above by*

$$|\mathcal{N}_{A_j}(\alpha, I_j, 2^i)| \leq \exp\left(\gamma_1 \alpha^{-1} \cdot (2^{i/z} + 1) \cdot |I_j| \log |I_j| \cdot d_{\text{VC}} \cdot \log |\Omega_G|\right),$$

*for some universal constant $\gamma$.*

*Proof.* Consider any solution $\mathcal{S}$ consisting of at most $|I_j|$ centers serving points in $A_j$. We consider the range space $\mathcal{R} = \{R(\mathcal{S}, r) \mid r \in \mathbb{R}_{\geq 0}\}$ composed of ranges $R(\mathcal{S}, r) = \{p \in A_j \mid \text{cost}(p, \mathcal{S}) \geq r\}$. By assuming, for $|\mathcal{S}| = 1$, the VC dimension of $\mathcal{R}$ is $d_{VC}$. If $|\mathcal{S}| = |I_j|$, then we are considering the $|I_j|$-fold intersection of ranges in $R(\mathcal{S}, r)$, as $p \in R(\mathcal{S}, r) \Leftrightarrow \forall c \in \mathcal{S} : \text{cost}(p, c) \geq r$. Then, the VC dimension of the $|I_j|$-fold intersection of ranges of metric balls in $\mathcal{R}$ (the set system defined by $R(\mathcal{S}, r)$) has VC dimension at most $O(|I_j| \log |I_j| \cdot d_{\text{VC}})$ [7, 27].

We now use ranges of the form $R(\mathcal{S}, r_t)$ to build a clustering net by discretizing the range from the smallest point-center distance in this type to the largest. To this end, let us consider $r_0 = 0$ and subsequently choose increments $\alpha\Lambda$, where we set

$$\Lambda := (2^{i(1-1/z)} + 1) \cdot \min_{p \in A_j} \text{dist}^z(p, \mathcal{A}).$$

We consider the cost vector ranges given by $r_i = r_{i-1} + \alpha\Lambda$. We denote by $t$ the final index and use $r_t = r_0 + t\alpha\Lambda = t\alpha\Lambda$ as an estimate for $\text{cost}(p, \mathcal{S})$. We would like to understand how large the final index $t$ can be. That is, it is sufficient that we solve

$$t\alpha\Lambda \leq \text{cost}(p, \mathcal{A}) + \text{cost}(p, \mathcal{S}), \ \forall p \in A_j,$$

For any two points $p, p' \in A_j$, we have $\text{cost}(p, \mathcal{A}) \leq 4\text{cost}(p', \mathcal{A})$ by definition of the groups and layers. Further, we likewise have, if $p'$ induced $q_p$ by Lemma 3.2 $\text{cost}(p, \mathcal{S}) \leq 2^{z-1} \cdot \text{cost}(p', \mathcal{S}) + 2^{z-1}\text{cost}(p, p') \leq 2^{z-1} \cdot 2^{i+1}\text{cost}(p', \mathcal{A}) + 2^{2z}\text{cost}(p', \mathcal{A})$. In other words, all points in $A_j$ satisfy $\text{cost}(p, \mathcal{S}) \leq \gamma \cdot \text{cost}(p, \mathcal{S})$ for some absolute constant $\gamma$ that depends on $z$.

By our choice of $\Lambda$, we can rewrite the inequality we would like to establish as

$$t\alpha \cdot (2^{i(1-1/z)} + 1) \cdot \min_{p \in A_j} \text{dist}^z(p, \mathcal{A}) \leq \gamma \cdot \text{cost}(p, \mathcal{S}), \ \forall p \in A_j.$$

Let $\gamma'$ be an absolute constant. If $2^{i(1-1/z)} \leq 1$, then $t = \gamma' \cdot \alpha^{-1}$ is a valid upper bound. If $2^{i(1-1/z)} \geq 1$, then $t = \gamma' \cdot \alpha^{-1} \cdot 2^{i/z}$ is a valid upper bound. The number of elements in this sequence is thus at most $\gamma' \cdot \alpha^{-1} \cdot (2^{i/z} + 1)$.

We build a cost vector $v'$ such that the cost vector $v^{\mathcal{S}}$ induced by $\mathcal{S}$ on the points in $A_j$ satisfies the clustering net property. For this, we set $v'_p = r_i$ if $p \in R(\mathcal{S}, r_i) \setminus R(\mathcal{S}, r_{i+1})$. This implies $r_i \leq v_p^{\mathcal{S}} \leq r_{i+1}$, and we have

$$|v'_p - v_p^{\mathcal{S}}| \leq r_{i+1} - r_i \leq \alpha\Lambda \leq \alpha \cdot (\text{dist}^{z-1}(p, \mathcal{S}) \cdot \text{dist}(p, \mathcal{A}) + \text{dist}^z(p, \mathcal{A})).$$

That is, the cost vector $v'$ has the error bound as required by Definition 3.1. We now bound the number of cost vectors $v'$ we construct in this manner. Since the VC dimension of the set system $R(\mathcal{S}, r)$ is at most $O(|I_j| \log |I_j| d_{\text{VC}})$, the Sauer-Shelah lemma [14] implies that the number of sets is at most $\exp\left(O(|I_j| \log |I_j| d_{\text{VC}} \log |\Omega_G|)\right)$. To construct $v'$, we select a different set $R(\mathcal{S}, r_i)$ for each $i \in \{0, \ldots t\}$. Thus taking the set of all $t$-combinations of the $R(\mathcal{S}, r_i)$ is an upper bound for the number of constructed cost vectors. Each cost vector can potentially be induced by its own solution, which we therefore add to the clustering net. Therefore the clustering net has size at most

$$\exp\left(\gamma' \alpha^{-1} (2^{i/z} + 1) \cdot |I_j| \log |I_j| \cdot d_{\text{VC}} \cdot \log |\Omega_G|\right),$$

as desired. ☐

**Combining Clustering Nets for Clustering Unions** We now consider a fixed interaction profile and prove the following claim:

**Claim 3.8** (Combining Approximate Solutions for Main Groups)**.** *Let $G$ be a main group with a sample $\Omega_G$ and let $\mathcal{S}$ be the (potentially infinite) set of solutions with fixed interaction profile $I$. Suppose $\alpha \geq \varepsilon^{2z+2}/k$ and $i \leq \log(\gamma\varepsilon^{-z})$. For all $g \in [k]$, there exists a type $i$ clustering net $\mathcal{N}_{U_{g+1}}(\alpha, \sum_{j \leq g} I_j, 2^i)$ of size*

$$\left|\mathcal{N}_{U_{g+1}}\left(\alpha, \sum_{j \leq g} I_j, 2^i\right)\right| \leq \exp\left(\gamma_2 \alpha^{-1} \cdot (2^{i/z} + 1) \cdot \sum_{j \leq g} |I_j| \log |I_j| \cdot d_{\text{VC}} \cdot \log(k|\Omega_G|)\right),$$

*for some universal constant $\gamma_2$.*

*Proof.* Consider the $g$-th cluster union in sorted order $A_g$, with its own clustering net $\mathcal{N}_{A_g}(\alpha, I_g, 2^i)$, and all the cheaper cluster unions $A_j$ with $j \leq g-1$. We are going to show that the following construction is a valid clustering net for $U_g$. For each $j \leq g$, points $p \in A_j$ can either be served by the $I_j$ centroids belonging to the clustering net $\mathcal{N}_{A_j}(\alpha, I_j, 2^i)$ without considering centroids in $\bigcup_{j' \leq j-1} \mathcal{N}_{A_{j'}}(\alpha, I_{j'}, 2^i)$, or by centroids belonging to clustering nets of clusters with lower average cost, i.e., centroids in $\bigcup_{j' \leq j-1} \mathcal{N}_{A_{j'}}(\alpha, I_{j'}, 2^i)$.

**Base Case:** $j = 1$**.** This is an immediate consequence of Claim 3.7.

**Extended Case:** $j > 1$**.** We consider a solution $\mathcal{S}$ consistent with the given interaction profile $I$. We split the clusters serving $A_{g+1}$ into the centers that are not close to any point in $U_g$, denoted by $\mathcal{S}_{g+1}$ and the centers that are close to some point in $U_g$, denoted by $\mathcal{S}_g$. For the former, we obtain a clustering net $\mathcal{N}_{A_{g+1}}(\alpha, I_{g+1}, 2^i)$ per Claim 3.7. We select a solution $\hat{\mathcal{S}}_{g+1}$ from this net. For the latter, we have $\sum_{j \leq g} I_j$ centers, which we assign to the clusters in $\cup_{j \leq g} A_j$. There are $\exp\left(\sum_{j \leq g} |I_j| \cdot \log k\right)$ such assignments. For any assignment, we place a center on the center of the cluster of the corresponding ring in $\cup_{j \leq g} A_j$, obtaining a solution $\hat{\mathcal{S}}_g$.

$\hat{\mathcal{S}}_{g+1}$ satisfies the clustering net guarantee for $\mathcal{S}_{g+1}$. For $\mathcal{S}_g$, we argue this as follows. Let $A'$ be a ring in $\cup_{j \leq g} A_j$ and let $\Delta'$ be the average cost of $A'$. Consider any center $c \in \mathcal{S}_g$ and assign $c$ to the clusters $A'$ if, $a'$ being the center of $A'$, we have $\text{cost}(c, a') \leq \gamma \cdot 2^i \cdot \Delta'$. We first note that for a sufficient large choice of the constant $\gamma$, this has to hold for some $a'$ as otherwise $c$ would not be counted among $I(\mathcal{S})_g$. This in turn implies that $\text{cost}(c, a') \lesssim 2^i \cdot \phi^{-1} \cdot \text{cost}(p, \mathcal{A}) \lesssim \phi^{-1} \cdot \text{cost}(p, \mathcal{S})$, due to the layering property. We argue that for any point $p \in A_{g+1}$, $\text{cost}(p, a')$ is very close to $\text{cost}(p, c)$. By Lemma 3.2, we have

$$|\text{cost}(p, a') - \text{cost}(p, c)| \leq \beta \cdot \text{cost}(p, c) + \left(\frac{\beta + 2z}{\beta}\right)^{z-1} \cdot \text{cost}(c, a')$$

$$\lesssim \beta \cdot \text{cost}(p,c) + \left(\frac{\beta+2z}{\beta}\right)^{z-1} \cdot \phi^{-1} \cdot \text{cost}(p,\mathcal{S})$$

We compose the solution $\hat{\mathcal{S}}_g$ with solution $\hat{\mathcal{S}}_{g+1}$. Taking the minimum over all center assignments, we have

$$\begin{aligned}\left|\text{cost}(p,\mathcal{S}_{g+1}) - \text{cost}(p,\hat{\mathcal{S}}_i)\right| &\lesssim \alpha \cdot (\text{dist}^{z-1}(p,\mathcal{S}) \cdot \text{dist}(p,\mathcal{A}) + \text{dist}^z(p,\mathcal{A})) \\ &\quad + \beta \cdot \text{cost}(p,\mathcal{S}) + \left(\frac{\beta+2z}{\beta}\right)^{z-1} \cdot \phi^{-1} \cdot \text{cost}(p,\mathcal{S})\end{aligned}$$

We now choose $\beta \lesssim \begin{cases} \alpha \cdot 2^{-i/z} & \text{if } 2^{i/z} \geq 1 \\ \alpha & \text{else} \end{cases}$. With Lemma 3.3 this implies

$$\beta \cdot \text{cost}(p,\mathcal{S}) \lesssim \alpha \cdot (\text{dist}^{z-1}(p,\mathcal{S}) \cdot \text{dist}(p,\mathcal{S}) + \text{dist}^z(p,\mathcal{A})).$$

To bound the final term, we have by our choice of $\beta$, $\phi$ (and for an appropriate choice of absolute constants) and again using Lemma 3.3 and $\alpha \geq \varepsilon^{2z+2}/k$

$$\begin{aligned}\left(\frac{\beta+2z}{\beta}\right)^{z-1} \cdot \gamma \cdot \phi^{-1}/2 \cdot \text{cost}(p,\mathcal{S}) &\lesssim \left(\frac{1+2^{i/z}}{\alpha}\right)^{z-1} \cdot \phi^{-1} \cdot \text{cost}(p,\mathcal{S}) \\ &\lesssim (\alpha \cdot \varepsilon)^{z-1} \cdot \phi^{-1} \cdot \text{cost}(p,\mathcal{S}) \\ &\lesssim (\alpha \cdot \varepsilon)^{z-1} \cdot \varepsilon^{10z}\text{cost}(p,\mathcal{S}) \\ &\lesssim \alpha \cdot (\text{dist}^{z-1}(p,\mathcal{S})\text{dist}(p,\mathcal{A}) + \text{dist}^z(p,\mathcal{A}))\end{aligned}$$

Collecting all of the terms then yields

$$\left|\text{cost}(p,\mathcal{S}_{g+1}) - \text{cost}(p,\hat{\mathcal{S}}_i)\right| \leq 3\alpha \cdot (\text{dist}^{z-1}(p,\mathcal{S}) \cdot \text{dist}(p,\mathcal{A}) + \text{dist}^z(p,\mathcal{A}))$$

which is the desired clustering net guarantee up to an appropriate rescaling of $\alpha$. The size of the constructed type $i$ clustering net $\mathcal{N}_{U_{g+1}}(\alpha, \sum_{j\leq g} I_j, 2^i)$ is

$$\left|\mathcal{N}_{A_{g+1}}(\alpha, I_{g+1}, 2^i)\right| \cdot k^{\sum_{j\leq g}|I_j|} \leq \exp\left(\gamma_2\alpha^{-1} \cdot (2^{i/z}+1) \cdot \sum_{j\leq g+1} |I_j| \log |I_j| \cdot d_{\text{VC}} \cdot \log(k|\Omega_G|)\right),$$

for some constant $\gamma_2$. □

We now finish the proof of Lemma 3.6.

*Proof of Lemma 3.6.* By Claim 3.8, we know that, for each interaction profile, there exists a type $i$ clustering net of size at most

$$\exp\left(\gamma_2\alpha^{-1} \cdot (2^{i/z}+1) \cdot \sum_{g\in[k]} |I_j| \log |I_j| \cdot d_{\text{VC}} \cdot \log(k|\Omega_G|)\right),$$

for an appropriate constant $\gamma_2 > 0$. Additionally, by Observation 3.5, we know that the number of possible interaction profiles is upper bounded by $(2e)^k$. Hence, the overall resulting clustering net has size at most

$$(2e)^k \cdot \exp\left(\gamma_2\alpha^{-1} \cdot (2^{i/z}+1) \cdot \sum_{g\in[k]} |I_j| \log |I_j| \cdot d_{\text{VC}} \cdot \log(k|\Omega_G|)\right),$$

which, by taking logarithms and recalling that $\sum_{i\in[k]} I_i \leq k$, yields a type $i$ clustering net of size at most

$$\exp\left(\gamma\alpha^{-1} \cdot (2^{i/z}+1) \cdot k \log k \cdot d_{\text{VC}} \cdot \log(k|\Omega_G|)\right),$$

for an appropriate constant $\gamma > 0$. This concludes the proof. □

### 3.2 Clustering Nets for Outer Groups

We now consider the clusters composing an outer group $G^O_{b,\ell}$, or simply $G$. The goal of this section is to show a corresponding result, in the case of outer groups, to Lemma 3.6:

**Lemma 3.9.** *For any outer group $G$, a sample $\Omega_G$ and any $\alpha \geq \varepsilon^{2z+2}/k$, there exists a type $2z$ clustering net $\mathcal{N}_G(\alpha, k, 2^{2i})$ for $G$ of size*

$$\left|\mathcal{N}_G(\alpha, k, 2^{2z})\right| \leq \exp\left(\gamma \cdot \alpha^{-1} \cdot k \log k \cdot d_{\mathrm{VC}} \cdot \log(k|\Omega_G|)\right),$$

*for some universal constant $\gamma > 0$.*

The techniques to prove such a result are similar to the ones used in Section 3.1, with some crucial differences arising from the fact that outer groups are arranged in a single outer ring.

Specifically, in the clustering net construction, we consider, for each cluster, the innermost ring of that cluster's outer rings, falling in the layer the group belongs to. We then aggregate all clusters whose innermost ring out of the outer rings has (roughly) the same cost, and sort all these cluster unions by such cost. Akin to Section 3.1, we let a more expensive cluster union be served by its own clustering net or by cheaper cluster unions' clustering nets. Formally, cluster $C_g \cap \Omega_G$'s innermost of the outer rings has a cost of about $2^{j_{g,\ell}}\Delta_{C_g}$, where $j_{g,\ell}$ denotes the exponent of cluster $C_g \cap \Omega_G$'s innermost of the outer rings, thus falling in the $\ell$-th layer, to which group $G$ belongs. Let $\Delta^O_1 = \min_{g\in[k]} 2^{j_{g,\ell}}\Delta_{C_g}$ and $\Delta^O_{k'} = \max_{g\in[k]} 2^{j_{g,\ell}}\Delta_{C_g}$. We also let $O_1$ be the set composed of the union of all clusters with $\Delta^O_g \leq 4\Delta^O_1$. We recursively define $U^O_j = \cup_{j'=1}^{j-1} O_{j'}$, with $U^O_1 = \emptyset$ and $O_j = G \setminus U^O_j$. That is, $O_2$ is the set composed of the clusters $C_{g'}$ with $\Delta^O_{g'} \leq 4 \cdot \min_{C_g \in G\setminus O_1} \Delta^O_g$. This is illustrated in Figure 2 and formalized in the proof of Claim 3.10.

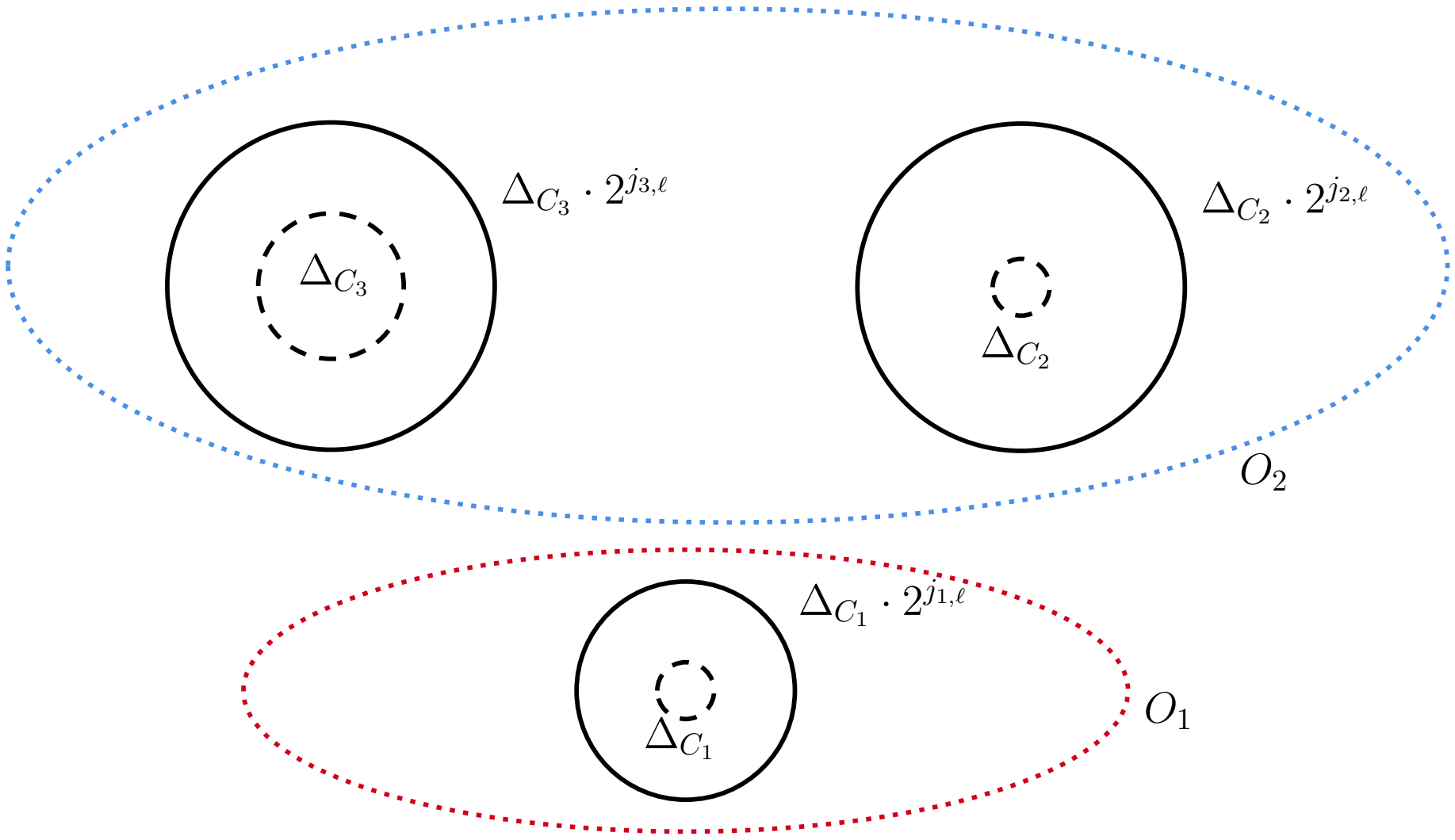

Figure 2: Black dashed circles depict the cluster core along with its average cost. Black solid circles instead represent the cluster's innermost of the outer rings falling the $\ell$-th layer, to which group $G$ belongs. As we may observe, even though the average cost may be different (see $\Delta_{C_2}$ vs. $\Delta_{C_3}$), they still belong to the same cluster union $O_2$ (blue dotted oval) because the $\ell$-th layering induces an innermost of the outer rings that, in the case of $C_2 \cap G$, is much further than the innermost of the outer rings for $C_3 \cap G$. Finally, points in $O_2$ are only allowed to be served by centroids in its own approximate centroid set or in $O_1$'s (red dotted oval) approximate centroid set.

In order to be able to prove Lemma 3.9, another important aspect that we need to have a handle over is the number of interaction profiles. If we were to consider all outer rings, the number of interaction profiles would increase significantly. This is in contrast to Section 3.1, where the number of main rings is at most $O(z\log(z\alpha^{-1}))$. However, since we only consider the innermost ring and sort according to its cost, there are at most $k$ such rings and, thus, the number of interaction profiles is still at most $|\Omega_G|^k$ by Observation 3.5. This is formalized in the proof of Lemma 3.9.

Hence, we again consider a fixed interaction profile and first prove the following claim:

**Claim 3.10** (Combining Approximate Solutions for Outer Groups)**.** *Let $G$ be an outer group with a sample $\Omega_G$ and let $\mathcal{S}$ be the (potentially infinite) set of solutions with fixed interaction profile $I$. Suppose $\alpha \geq \varepsilon^{2z+2}/k$ For all $g \in [k]$, there exists a type $2z$ clustering net $\mathcal{N}_{U_g^O}(\alpha, \sum_{j \leq g} I_j, 2^{2z})$ of size*

$$\left|\mathcal{N}_{U_g^O}\left(\alpha, \sum_{j\leq g} I_j, 2^{2z}\right)\right| \leq \exp\left(\gamma \cdot \alpha^{-1} \cdot \sum_{j \leq g} I_j \log I_j \cdot d_{\text{VC}} \cdot \log(k|\Omega_G|)\right),$$

*for some universal constant $\gamma$.*

*Proof.* Consider the $i$-th cluster union in sorted order $O_i$, with its own clustering net $\mathcal{N}_{O_i}(\alpha, I_g, 8)$, and all the cheaper cluster unions $O_j$ with $j \leq g-1$. We are going to show that the following construction is a valid clustering net for $U_g^O$. For each $j \leq g$, points $p \in O_j$ can either be served by the $I_j$ centers belonging to the clustering net $\mathcal{N}_{O_j}(\alpha, I_j, 8)$ without considering centers in $\bigcup_{j' \leq j-1} \mathcal{N}_{O_{j'}}(\alpha, I_{j'}, r)$, or by centers belonging to clustering nets of clusters with lower average cost, i.e., centers in $\bigcup_{j' \leq j-1} \mathcal{N}_{O_{j'}}(\alpha, I_{j'}, r)$.

**Base Case:** $O_1$**.** This is an immediate consequence of Claim 3.7.

**Extended Case:** $U_g^O$ **and** $O_{g+1}$**.** We consider a solution $\mathcal{S}$ consistent with the given interaction profile $I$. We split the clusters serving $O_{g+1}$ into the centers $c \subset \mathcal{S}$ that are close to some point in $U_g^O$, meaning there exists a point $p \in U_g^O$ such that $\text{cost}(p, c) \leq \gamma \cdot 2^{2z} \cdot \text{cost}(p, \mathcal{A})$ as per assumption on the interaction profile and a sufficiently large constant $\gamma$, denoted by $\mathcal{S}_g$ and the remaining centers that are not close to any point in $U_g^O$ but satisfy $\text{cost}(p, c) \leq \gamma \cdot 2^{2z} \cdot \text{cost}(p, \mathcal{A})$, for some point $p \in O_{g+1}$, denoted by $\mathcal{S}_{g+1}$. The remaining centers of $\mathcal{S}$ cannot serve an point $p \in O_{g+1}$ with cost $\gamma \cdot 2^{2z}\text{cost}(p, \mathcal{A})$ and we therefore do not have to consider them.

For $\mathcal{S}_{g+1}$, we obtain a clustering net $\mathcal{N}_{O_{g+1}}(\alpha, I_{g+1}, 2^{2z})$ per Claim 3.7, yielding a solution $\hat{S}_{g+1}$. For the latter, we snap any $c \in \mathcal{S}_g$ to the closest point in $U_g^O$, obtaining a solution $\hat{\mathcal{S}}_g$. The number of solutions $\hat{\mathcal{S}}_g$ generated in this way may be upper bounded by $\binom{|\Omega_G|}{\sum_{j\leq g}|I_j|} \leq \exp\left(\sum_{j \leq g} |I_j| \log |\Omega_G|\right)$. $\hat{\mathcal{S}}_{g+1}$ satisfies the clustering net guarantee for $\mathcal{S}_{g+1}$. When comparing the cost for any center $c \in \mathcal{S}_g$ and the corresponding center $c' \in \hat{\mathcal{S}}_g$, we argue this as follows. Let $p$ be any point in $O_{g+1}$. We then have

$$\text{cost}(c, \tilde{c}) \lesssim \phi^{-1}\text{cost}(p, \mathcal{A}).$$

By Lemma 3.2, we obtain

$$\begin{aligned}|\text{cost}(p, c) - \text{cost}(p, \tilde{c})| &\leq \beta \cdot \text{cost}(p, c) + \left(1 + \frac{2z}{\beta}\right)^{z-1} \text{cost}(c, \tilde{c}) \\ &\lesssim \beta \cdot \text{cost}(p, c) + \left(1 + \frac{2z}{\beta}\right)^{z-1} \phi^{-1}\text{cost}(p, \mathcal{A}),\end{aligned}$$

for all $c, \tilde{c}$. Thus, since the above is true for all centers, then it must be true for the minimum-cost assignment centers of the centers in $\hat{\mathcal{S}}_g$. Further adding the centers from $\hat{\mathcal{S}}_{g+1}$, we obtain for any type $2z$ point $p \in O_{g+1}$

$$\left|\text{cost}(p, \mathcal{S}) - \text{cost}(p, \hat{\mathcal{S}}_g \cup \hat{\mathcal{S}}_{g+1})\right| \lesssim \alpha \cdot \text{cost}(p, \mathcal{S}) + \beta \cdot \text{cost}(p, \mathcal{S}) + \left(1 + \frac{2z}{\beta}\right)^{z-1} \phi^{-1}\text{cost}(p, \mathcal{A})$$

We choose $\beta = \alpha$, which controls the second term on the RHS. For the remaining term, we then have

$$\left(1 + \frac{2z}{\beta}\right)^{z-1} \phi^{-1} \lesssim \alpha^{z-1} \cdot \left(\frac{\varepsilon}{k}\right)^{10z} \lesssim \alpha$$

as $\alpha \geq \varepsilon^{2z+2}/k$. The size of the constructed type $2z$ clustering net $\mathcal{N}_{U_{g+1}}(\alpha, \sum_{j\leq g} I_j, 2^{2z})$ is

$$\left|\mathcal{N}_{U_{g+1}}\left(\alpha, \sum_{j \leq g} I_j, 2^{2z}\right)\right| \cdot \exp\left(\sum_{j \leq g} |I_j| \log |\Omega_G|\right) \leq \exp\left(\gamma\alpha^{-1} \cdot \sum_{j \leq g} |I_j| \log |I_j| \cdot d_{\text{VC}} \cdot \log |\Omega_G|\right),$$

for some constant $\gamma$. ∎

We now finish the proof of Lemma 3.9.

*Proof of Lemma 3.9.* By Claim 3.10, we know that, for each fixed interaction profile, there exists a type $2z$ clustering net of size at most

$$\exp\left(\gamma\alpha^{-1}\log\left(\alpha^{-1}\right)\cdot\sum_{g\in[k]}|I_j|\log|I_j|\cdot d_{\mathrm{VC}}\cdot\log|G|\right),$$

for an appropriate absolute constant $\gamma$. By Observation 3.5, the number of possible interaction profiles is upper bounded by $(2e|\Omega_G|)^k$. Hence, the overall resulting clustering net has size at most

$$(2e|\Omega_G|)^k\cdot\exp\left(\gamma\cdot\alpha^{-1}\cdot\sum_{g\in[k]}|I_j|\log|I_j|\cdot d_{\mathrm{VC}}\cdot\log|\Omega_G|\right)\le\exp\left(2\gamma\alpha^{-1}\cdot k\log k\cdot d_{\mathrm{VC}}\cdot\log|G|\right).$$

□

# 4 Concentration Bounds

The purpose of this section is to use the nets obtained in the previous sections to prove Lemmas 2.3, 2.4, 2.5, 2.6, and 2.7 for the $(k,z)$-clustering task.

For the remainder of this paper, we only discuss the $v$ cost vectors as corresponding to clusters of a specific type. For example, in this section we use $v\in\mathbb{R}^{n_i}$ to represent the cost vector with $v_j=\mathrm{cost}(p_j,\mathcal{S})$ for $p_j\in\bigcup_{C\in T_i}C$, where $n_i=\left|\bigcup_{C\in T_i}C\right|$ is the number of points belonging to clusters in type $i$.

In addition to standard concentration inequalities, we will also sometimes employ a reduction to Gaussian processes by way of the following lemma.

**Lemma 4.1** (Appendix B.3 of [45]). *Let $g_p$ be independent standard Gaussian random variables. Then,*

$$\mathbb{E}_\Omega\left[\sup_{\mathcal{S}}\left|\frac{\frac{1}{|\Omega|}\left[\left(\sum_{C_j\in T_i}\sum_{p\in C_j\cap\Omega}\mathrm{cost}(p,\mathcal{S})\cdot w_p\right)-\|v^{\mathcal{S}}\|_1\right]}{\mathrm{cost}(G,\mathcal{S})+\mathrm{cost}(G,\mathcal{A})}\right|\right]$$
$$\le\sqrt{2\pi}\mathbb{E}_\Omega\mathbb{E}_g\left[\sup_{\mathcal{S}}\left|\frac{\frac{1}{|\Omega|}\sum_{C_j\in T_i}\sum_{p\in C_j\cap\Omega}\mathrm{cost}(p,\mathcal{S})\cdot w_p\cdot g_p}{\mathrm{cost}(G,\mathcal{S})+\mathrm{cost}(G,\mathcal{A})}\right|\right].$$

A useful inequality for controlling Gaussian processes is given in the following statement.

**Lemma 4.2** (Lemma 2.3 of [43]). *Let $g_i\sim\mathcal{N}(0,\sigma_i^2)$, $i\in[n]$ be Gaussian random variables and suppose $\sigma_i\le\sigma$ for all $i$. Then $\mathbb{E}[\max_{i\in[n]}|g_i|]\le2\sigma\cdot\sqrt{2\ln n}$.*

We will also require the following events for main groups.

**Definition 4.3.** *Let $\mathcal{E}$ denote the event that $\frac{1}{|\Omega|}\sum_{p\in C\cap\Omega}w_p=(1\pm\varepsilon)\cdot|C|$ for all clusters $C$.*

**Lemma 4.4** (Lemma 19 of [21]). *Let $\Omega$ be any sample obtained for a group $G$ using Algorithm 1. Then event $\mathcal{E}$ holds with probability $1-k\cdot\exp\left(-\frac{\varepsilon^2}{9k}|\Omega|\right)$.*

This has the following immediate implications:

**Corollary 4.5.** *Let $|\Omega|\ge\gamma\cdot k\varepsilon^{-2}\log\left(k\varepsilon^{-1}\right)$ and suppose $2^i\le\gamma'\cdot\varepsilon^{-z}$ for an absolute constants $\gamma$ and $\gamma'$. Then event $\mathcal{E}$ holds for all clusters $C_j$ with probability at least $1-\varepsilon\cdot2^{-2i}/k$.*

**Fact 4.6.** *If event $\mathcal{E}$ holds, then the number of points in $\Omega$ sampled from clusters of type $i$ is given by $|T_i\cap C_j\cap\Omega|\le4\cdot\frac{k_i\cdot|\Omega|}{k}$, where $k_i$ is the number of clusters of type $i$.*

### 4.1 Proof of Lemma 2.3

We first apply the symmetrization, so we wish to show

$$\mathbb{E}_\Omega \mathbb{E}_g \left[ \sup_{\mathcal{S}} \left| \frac{\frac{1}{|\Omega|} \sum_{C_j \in T_{small}} \sum_{p \in C_j \cap \Omega} \text{cost}(p, \mathcal{S}) \cdot w_p \cdot g_p}{\text{cost}(G, \mathcal{S}) + \text{cost}(G, \mathcal{A})} \right| \right] \leq \frac{\varepsilon}{\log^4(k\varepsilon^{-1})}.$$

We first write $\text{cost}(p, \mathcal{S})$ as a telescoping series. Specifically,

$$\text{cost}(p, \mathcal{S}) = \underbrace{v_p^{\mathcal{S}} - v_p^{\mathcal{S}, h_{\max}}}_{\text{Term 1}} + \underbrace{\sum_{h=0}^{h_{\max}-1} v_p^{\mathcal{S}, h+1} - v_p^{\mathcal{S}, h}}_{\text{Term 2}},$$

where we determine $v_p^{\mathcal{S},h}$ as the value of $p$ drawn from a solution from the clustering net $\mathcal{N}_G(2^{-h}, k, 2^{2z})$ approximating $\mathcal{S}$, $2^{-h_{\max}} = \varepsilon^{2z+2}/k$, that is the maximum value of $\alpha$ in the clustering nets. We will control the Gaussian process for each term individually, that is, we write

$$\begin{aligned}
&\mathbb{E}_\Omega \mathbb{E}_g \left[ \sup_{\mathcal{S}} \left| \frac{\frac{1}{|\Omega|} \sum_{C_j \in T_{small}} \sum_{p \in C_j \cap \Omega} \text{cost}(p, \mathcal{S}) \cdot w_p \cdot g_p}{\text{cost}(G, \mathcal{S}) + \text{cost}(G, \mathcal{A})} \right| \right] \\
\leq & \mathbb{E}_\Omega \mathbb{E}_g \left[ \sup_{\mathcal{S}} \left| \frac{\frac{1}{|\Omega|} \sum_{C_j \in T_{small}} \sum_{p \in C_j \cap \Omega} (v_p^{\mathcal{S}} - v_p^{\mathcal{S}, h_{max}}) \cdot w_p \cdot g_p}{\text{cost}(G, \mathcal{S}) + \text{cost}(G, \mathcal{A})} \right| \right] \\
&+ \mathbb{E}_\Omega \mathbb{E}_g \sum_{h=1}^{h_{\max}-1} \left[ \sup_{\mathcal{S}} \left| \frac{\frac{1}{|\Omega|} \sum_{C_j \in T_{small}} \sum_{p \in C_j \cap \Omega} (v_p^{\mathcal{S}, h+1} - v_p^{\mathcal{S}, h}) \cdot w_p \cdot g_p}{\text{cost}(G, \mathcal{S}) + \text{cost}(G, \mathcal{A})} \right| \right]
\end{aligned}$$

Showing that both terms are upper bounded by $\frac{\varepsilon}{\log^4(k\varepsilon^{-1})}$ yields the claim.

**Term** 1 We apply the Cauchy Schwarz inequality to obtain

$$\begin{aligned}
&\mathbb{E}_\Omega \mathbb{E}_g \left[ \sup_{\mathcal{S}} \left| \frac{\frac{1}{|\Omega|} \sum_{C_j \in T_{small}} \sum_{p \in C_j \cap \Omega} (v_p^{\mathcal{S}} - v_p^{\mathcal{S}, h_{max}}) \cdot w_p \cdot g_p}{\text{cost}(G, \mathcal{S}) + \text{cost}(G, \mathcal{A})} \right| \right] \\
\leq & \frac{1}{|\Omega|} \mathbb{E}_\Omega \sqrt{\sum_{C_j \in T_{small}} \sum_{p \in C_j \cap \Omega} \left( \frac{(v_p^{\mathcal{S}} - v_p^{\mathcal{S}, h_{max}}) \cdot w_p}{\text{cost}(G, \mathcal{S}) + \text{cost}(G, \mathcal{A})} \right)^2} \cdot \mathbb{E}_g \sqrt{\sum_{p \in C_j \cap \Omega} g_p^2}
\end{aligned}$$

The second factor may be controlled by $\sqrt{|\Omega|}$ due to Jensen's inequality and noting that $\mathbb{E}_g \sum_{p \in C_j \cap \Omega} g_p^2 = |C_j \cap \Omega|$. For the first term, we know that $|v_p^{\mathcal{S}} - v_p^{\mathcal{S}, h_{max}}| \lesssim \varepsilon \cdot \varepsilon^{2z+2}/k \cdot v_p^{\mathcal{A}}$,since we are considering $T_{small}$. Therefore

$$\begin{aligned}
&\sqrt{\sum_{C_j \in T_{small}} \sum_{p \in C_j \cap \Omega} \left( \frac{(v_p^{\mathcal{S}} - v_p^{\mathcal{S}, h_{max}}) \cdot w_p}{\text{cost}(G, \mathcal{S}) + \text{cost}(G, \mathcal{A})} \right)^2} \\
\lesssim & \varepsilon^{2z+2}/k \cdot \sqrt{\sum_{C_j \in T_{small}} \sum_{p \in C_j \cap \Omega} \left( \frac{v_p^{\mathcal{A}} \cdot \frac{|C_i \cap G| \cdot \text{cost}(G, \mathcal{A})}{\text{cost}(C_i, \mathcal{A})}}{\text{cost}(G, \mathcal{S}) + \text{cost}(G, \mathcal{A})} \right)^2} \\
\lesssim & \varepsilon^{2z+2}/k \cdot \sqrt{\sum_{C_j \in T_{small}} \sum_{p \in C_j \cap \Omega} \left( \frac{\text{cost}(G, \mathcal{A})}{\text{cost}(G, \mathcal{S}) + \text{cost}(G, \mathcal{A})} \right)^2} \lesssim \varepsilon^{2z+2}/k \cdot \sqrt{|\Omega|}
\end{aligned}$$

Hence, we may conclude that Term 1 is bounded by at most $\varepsilon^{2z+2}/k \lesssim \frac{\varepsilon}{\log^4(k \cdot \varepsilon^{-1})}$.

**Term 2** We condition on $\Omega$. Now $\frac{\frac{1}{|\Omega|}\sum_{C_j\in T_{small}}\sum_{p\in C_j\cap\Omega}(v_p^{\mathcal{S},h+1}-v_p^{\mathcal{S},h})\cdot w_p\cdot g_p}{\text{cost}(G,\mathcal{S})+\text{cost}(G,\mathcal{A})}$ is Gaussian distributed with mean 0 and variance

$$\frac{1}{|\Omega|^2}\frac{\sum_{C_j\in T_{small}}\left(\sum_{p\in C_j\cap\Omega}(v_p^{\mathcal{S},h+1}-v_p^{\mathcal{S},h})\cdot w_p\right)^2}{\left(\text{cost}(G,\mathcal{S})+\text{cost}(G,\mathcal{A})\right)^2}$$
$$\lesssim\frac{2^{-2h}}{|\Omega|^2}\frac{\sum_{C_j\in T_{small}}\sum_{p\in C_j\cap\Omega}\left(v_p^{\mathcal{S}}\cdot\frac{|C_i\cap G|\cdot\text{cost}(G,\mathcal{A})}{\text{cost}(C_i,\mathcal{A})}\right)^2}{\left(\text{cost}(G,\mathcal{S})+\text{cost}(G,\mathcal{A})\right)^2}\lesssim\frac{2^{-2h}}{|\Omega|}.$$

Therefore, using Lemma 4.2, the bounds on the clustering nets for type $2z$ from Lemma 3.6 and the law of total expectation, we have

$$\mathbb{E}_\Omega\mathbb{E}_g\sum_{h=1}^{h_{\max}-1}\left[\sup_{\mathcal{S}}\left|\frac{\frac{1}{|\Omega|}\sum_{C_j\in T_{small}}\sum_{p\in C_j\cap\Omega}(v_p^{\mathcal{S},h+1}-v_p^{\mathcal{S},h})\cdot w_p\cdot g_p}{\text{cost}(G,\mathcal{S})+\text{cost}(G,\mathcal{A})}\right|\right]$$
$$\lesssim\sum_\Omega\mathbb{E}_g\sum_{h=1}^{h_{\max}-1}\left[\sup_{\mathcal{S}}\left|\frac{\frac{1}{|\Omega|}\sum_{C_j\in T_{small}}\sum_{p\in C_j\cap\Omega}2^{-h}v_p^{\mathcal{S}}\cdot w_p\cdot g_p}{\text{cost}(G,\mathcal{S})+\text{cost}(G,\mathcal{A})}\right|\middle|\Omega\right]\cdot\mathbb{P}[\Omega]$$
$$\lesssim\sum_\Omega\sum_{h=1}^{h_{\max}-1}\sqrt{\frac{2^{-2h}}{|\Omega|}\cdot\log\exp\left(\gamma 2^h\cdot k\log k\cdot d_{\text{VC}}\cdot\log|\Omega|\right)}\cdot\mathbb{P}[\Omega]$$
$$\lesssim\sum_\Omega\sum_{h=1}^{h_{\max}-1}\sqrt{\frac{2^{-h}}{|\Omega|}\cdot\gamma k\log k\cdot d_{\text{VC}}\cdot\log|\Omega|}\cdot\mathbb{P}[\Omega]\lesssim\frac{\varepsilon}{\log^4(k\varepsilon^{-1})}$$

where the final bound follows from our choice of $\Omega$ and $x\geq y\log x$ being true for $y\log y\lesssim x$.

## 4.2 Proof of Lemma 2.4

As with the proof of Lemma 2.3, we start by applying the symmetrization, that is we will show

$$\mathbb{E}_\Omega\mathbb{E}_g\left[\sup_{\mathcal{S}}\left|\frac{\frac{1}{|\Omega|}\sum_{C_j\in T_i}\sum_{p\in C_j\cap\Omega}\text{cost}(p,\mathcal{S})\cdot w_p\cdot g_p}{\text{cost}(G,\mathcal{S})+\text{cost}(G,\mathcal{A})}\right|\right]\leq\frac{\varepsilon}{\log^4(k\varepsilon^{-1})}.$$

Again, we will write $\text{cost}(p,\mathcal{S})$ as a telescoping series:

$$\text{cost}(p,\mathcal{S})=\underbrace{v_p^{\mathcal{S}}-v_p^{\mathcal{S},h_{\max}}}_{\text{Term 1}}+\underbrace{\sum_{h=0}^{h_{\max}-1}v_p^{\mathcal{S},h+1}-v_p^{\mathcal{S},h}}_{\text{Term 2}}+\underbrace{q_p\mathcal{S}}_{\text{Term 3}},$$

where $q_p^{\mathcal{S}}$ is understood from Definition 2.2, we use $v_p^{\mathcal{S},0}=q_p^{\mathcal{S}}$ and otherwise determine $v_p^{\mathcal{S},h}$ as the value of $p$ drawn from a solution from the clustering net $\mathcal{N}_G(2^{-h},k,2^{2z})$ approximating $\mathcal{S}$, $2^{-h_{\max}}=\varepsilon^{2z+2}/k$, that is the maximum value of $\alpha$ in the clustering nets. As above, we will control the Gaussian process for each term individually, that is, we write

$$\mathbb{E}_\Omega\mathbb{E}_g\left[\sup_{\mathcal{S}}\left|\frac{\frac{1}{|\Omega|}\sum_{C_j\in T_i}\sum_{p\in C_j\cap\Omega}\text{cost}(p,\mathcal{S})\cdot w_p\cdot g_p}{\text{cost}(G,\mathcal{S})+\text{cost}(G,\mathcal{A})}\right|\right]$$
$$\leq\mathbb{E}_\Omega\mathbb{E}_g\left[\sup_{\mathcal{S}}\left|\frac{\frac{1}{|\Omega|}\sum_{C_j\in T_i}\sum_{p\in C_j\cap\Omega}(v_p^{\mathcal{S}}-v_p^{\mathcal{S},h_{max}})\cdot w_p\cdot g_p}{\text{cost}(G,\mathcal{S})+\text{cost}(G,\mathcal{A})}\right|\right]$$
$$+\mathbb{E}_\Omega\mathbb{E}_g\sum_{h=0}^{h_{\max}-1}\left[\sup_{\mathcal{S}}\left|\frac{\frac{1}{|\Omega|}\sum_{C_j\in T_i}\sum_{p\in C_j\cap\Omega}(v_p^{\mathcal{S},h+1}-v_p^{\mathcal{S},h})\cdot w_p\cdot g_p}{\text{cost}(G,\mathcal{S})+\text{cost}(G,\mathcal{A})}\right|\right]$$

$$+ \mathbb{E}_\Omega \mathbb{E}_g \left[ \sup_{\mathcal{S}} \left| \frac{\frac{1}{|\Omega|} \sum_{C_j \in T_i} \sum_{p \in C_j \cap \Omega} q_p^{\mathcal{S}} \cdot w_p \cdot g_p}{\text{cost}(G, \mathcal{S}) + \text{cost}(G, \mathcal{A})} \right| \right].$$

Showing that each term is upper bounded by $\frac{\varepsilon}{\log^4(k\varepsilon^{-1})}$ yields the claim.

**Term** 1 We apply the Cauchy Schwarz inequality to obtain

$$\mathbb{E}_\Omega \mathbb{E}_g \left[ \sup_{\mathcal{S}} \left| \frac{\frac{1}{|\Omega|} \sum_{C_j \in T_i} \sum_{p \in C_j \cap \Omega} (v_p^{\mathcal{S}} - v_p^{\mathcal{S}, h_{max}}) \cdot w_p \cdot g_p}{\text{cost}(G, \mathcal{S}) + \text{cost}(G, \mathcal{A})} \right| \right]$$
$$\leq \frac{1}{|\Omega|} \mathbb{E}_\Omega \sqrt{\sum_{C_j \in T_i} \sum_{p \in C_j \cap \Omega} \left( \frac{(v_p^{\mathcal{S}} - v_p^{\mathcal{S}, h_{max}}) \cdot w_p}{\text{cost}(G, \mathcal{S}) + \text{cost}(G, \mathcal{A})} \right)^2} \cdot \mathbb{E}_g \sqrt{\sum_{p \in C_j \cap \Omega} g_p^2}$$

The second factor may be controlled by $\sqrt{|\Omega|}$ due to Jensen's inequality and noting that $\mathbb{E}_g \sum_{p \in C_j \cap \Omega} g_p^2 = |C_j \cap \Omega|$. For the first term, we know that $|v_p^{\mathcal{S}} - v_p^{\mathcal{S}, h_{max}}| \lesssim \varepsilon^{2z+2}/k \cdot v_p^{\mathcal{S}} \lesssim \varepsilon^{2z+2}/k \cdot \varepsilon^{-z} \cdot v_p^{\mathcal{A}} \lesssim \varepsilon^{z+2}/k \cdot v_p^{\mathcal{A}}$, as $i \leq \log(\gamma \varepsilon^{-z})$. Therefore, we have

$$\sqrt{\sum_{C_j \in T_i} \sum_{p \in C_j \cap \Omega} \left( \frac{(v_p^{\mathcal{S}} - v_p^{\mathcal{S}, h_{max}}) \cdot w_p}{\text{cost}(G, \mathcal{S}) + \text{cost}(G, \mathcal{A})} \right)^2}$$
$$\lesssim \varepsilon^{z+2}/k \cdot \sqrt{\sum_{C_j \in T_i} \sum_{p \in C_j \cap \Omega} \left( \frac{v_p^{\mathcal{A}} \cdot \frac{|C_i \cap G| \cdot \text{cost}(G, \mathcal{A})}{\text{cost}(C_i, \mathcal{A})}}{\text{cost}(G, \mathcal{S}) + \text{cost}(G, \mathcal{A})} \right)^2}$$
$$\lesssim \varepsilon^{2z+2} \cdot \sqrt{\sum_{C_j \in T_i} \sum_{p \in C_j \cap \Omega} \left( \frac{\text{cost}(G, \mathcal{A})}{\text{cost}(G, \mathcal{S}) + \text{cost}(G, \mathcal{A})} \right)^2} \lesssim \varepsilon^{2z+2}/k \cdot \sqrt{|\Omega|}$$

Hence, we may conclude that Term 1 is bounded by at most $\varepsilon^{z+2}/k \lesssim \frac{\varepsilon}{\log^4(k \cdot \varepsilon^{-1})}$.

**Term** 2 We condition on $\Omega$. Now $\frac{\frac{1}{|\Omega|} \sum_{C_j \in T_i} \sum_{p \in C_j \cap \Omega} (v_p^{\mathcal{S}, h+1} - v_p^{\mathcal{S}, h}) \cdot w_p \cdot g_p}{\text{cost}(G, \mathcal{S}) + \text{cost}(G, \mathcal{A})}$ is Gaussian distributed with mean 0. Due to Lemma 3.6 and Lemma 3.4, we have

$$\begin{aligned} |v_p^{\mathcal{S}, h+1} - v_p^{\mathcal{S}, h}| &\lesssim 2^{-h} \cdot (v_p^{\mathcal{S}} - q_p^{\mathcal{S}} + v_p^{\mathcal{A}}) \\ &\lesssim 2^{-h} \cdot \left( \text{dist}^{z-1}(p, \mathcal{S}) \cdot \text{dist}(p, \mathcal{A}) + \text{dist}^z(p, \mathcal{A}) \right) \\ &\lesssim 2^{-h} \cdot 2^{i(1-1/z)} \cdot v_p^{\mathcal{A}} \end{aligned} \tag{9}$$

Furthermore, we have

$$\text{cost}(G, \mathcal{S}) \gtrsim \sum_{C_j \in T_i} 2^i \cdot \text{cost}(C_j, \mathcal{A}) \gtrsim \frac{k_i \cdot 2^i}{k} \cdot \text{cost}(G, \mathcal{A})) \tag{10}$$

where $k_i$ is the number of clusters of $G$ with type $i$ in $\mathcal{S}$.

For the variance, conditioned on event $\mathcal{E}$, using Fact 4.6 and using Equations 9 and 10

$$\frac{1}{|\Omega|^2} \frac{\sum_{C_j \in T_i} \left( \sum_{p \in C_j \cap \Omega} (v_p^{\mathcal{S}, h+1} - v_p^{\mathcal{S}, h}) \cdot w_p \right)^2}{(\text{cost}(G, \mathcal{S}) + \text{cost}(G, \mathcal{A}))^2}$$
$$\lesssim \frac{2^{-2h}}{|\Omega|^2} \cdot \frac{\sum_{C_j \in T_{small}} \sum_{p \in C_j \cap \Omega} \left( 2^{i(1-1/z)} \cdot v_p^{\mathcal{A}} \cdot \frac{|C_i \cap G| \cdot \text{cost}(G, \mathcal{A})}{\text{cost}(C_i, \mathcal{A})} \right)^2}{\left( \frac{k_i \cdot 2^i}{k} + 1 \right)^2 \text{cost}^2(G, \mathcal{A})}$$

$$\begin{aligned}
&\lesssim \frac{2^{-2h}}{|\Omega|^2} \cdot 2^{2i(1-1/z)} \cdot \frac{\sum_{C_j \in T_{small}} \sum_{p \in C_j \cap \Omega} \text{cost}^2(G,\mathcal{A})}{\left(\frac{k_i \cdot 2^i}{k}+1\right)^2 \text{cost}^2(G,\mathcal{A})} \\
&\lesssim \frac{2^{-2h}}{|\Omega|} \cdot 2^{2i(1-1/z)} \cdot \frac{\frac{k_i}{k}}{\left(\frac{k_i \cdot 2^i}{k}+1\right)^2} \\
&\lesssim \frac{2^{-2h}}{|\Omega|} \cdot \min(2^{i(1-2/z)}, 2^{-2i/z} \cdot \frac{k}{k_i}) \\
&\lesssim \frac{2^{-2h}}{|\Omega|} \cdot \min(2^{i(1-2/z)}, \frac{k}{2^{i/z}}) \qquad (11)
\end{aligned}$$

If we do not condition on event $\mathcal{E}$, then using the same derivation except for the last two steps, we obtain the cruder bounds

$$\frac{1}{|\Omega|^2} \frac{\sum_{C_j \in T_i} \left(\sum_{p \in C_j \cap \Omega} (v_p^{\mathcal{S},h+1} - v_p^{\mathcal{S},h}) \cdot w_p\right)^2}{(\text{cost}(G,\mathcal{S}) + \text{cost}(G,\mathcal{A}))^2} \lesssim \frac{2^{-2h}}{|\Omega|} \cdot 2^{2i}. \qquad (12)$$

Using the law of total expectation, Lemma 4.2, the bounds on the clustering nets for type $2z$ from Lemma 3.6, Equations 10 and 12, we have

$$\begin{aligned}
&\mathbb{E}_\Omega \mathbb{E}_g \sum_{h=1}^{h_{\max}-1} \left[\sup_{\mathcal{S}} \left| \frac{\frac{1}{|\Omega|} \sum_{C_j \in T_i} \sum_{p \in C_j \cap \Omega} (v_p^{\mathcal{S},h+1} - v_p^{\mathcal{S},h}) \cdot w_p \cdot g_p}{\text{cost}(G,\mathcal{S}) + \text{cost}(G,\mathcal{A})} \right|\right] \\
&\lesssim \sum_\Omega \left( \mathbb{E}_g \sum_{h=1}^{h_{\max}-1} \left[\sup_{\mathcal{S}} \left| \frac{\frac{1}{|\Omega|} \sum_{C_j \in T_i} \sum_{p \in C_j \cap \Omega} 2^{-h} v_p^{\mathcal{S}} \cdot w_p \cdot g_p}{\text{cost}(G,\mathcal{S}) + \text{cost}(G,\mathcal{A})} \right| \,\middle|\, \Omega, \mathcal{E} \right] \cdot \mathbb{P}[\Omega \wedge \mathcal{E}] \right. \\
&\quad \left. + \mathbb{E}_g \sum_{h=1}^{h_{\max}-1} \left[\sup_{\mathcal{S}} \left| \frac{\frac{1}{|\Omega|} \sum_{C_j \in T_i} \sum_{p \in C_j \cap \Omega} 2^{-h} v_p^{\mathcal{S}} \cdot w_p \cdot g_p}{\text{cost}(G,\mathcal{S}) + \text{cost}(G,\mathcal{A})} \right| \,\middle|\, \Omega, \overline{\mathcal{E}} \right] \cdot \mathbb{P}[\Omega \wedge \overline{\mathcal{E}}] \right). \\
&\lesssim \sum_\Omega \left( \sum_{h=1}^{h_{\max}-1} \sqrt{\frac{2^{-2h}}{|\Omega|} \cdot \min\left(2^{i(1-2/z)}, \frac{k}{2^{i/z}}\right) \log \exp\left(\gamma \cdot 2^h \cdot k \log k \cdot 2^{i/z} \cdot d_{\text{VC}} \cdot \log(k|\Omega|)\right)} \cdot \mathbb{P}[\Omega \wedge \mathcal{E}] \right. \\
&\quad \left. + \sum_{h=1}^{h_{\max}-1} \sqrt{\frac{2^{-2h}}{|\Omega|} \cdot 2^{2i} \log \exp\left(\gamma \cdot 2^h \cdot k \log k \cdot 2^{i/z} \cdot d_{\text{VC}} \cdot \log(k|\Omega|)\right)} \cdot \mathbb{P}[\Omega \wedge \overline{\mathcal{E}}] \right) \\
&\lesssim \sum_\Omega \left( \sqrt{\frac{1}{|\Omega|} \cdot \min\left(2^{i(1-1/z)}, k\right) \cdot \gamma \cdot k \log k \cdot d_{\text{VC}} \cdot \log(k|\Omega|)} \cdot \mathbb{P}[\Omega \wedge \mathcal{E}] \right. \\
&\quad \left. + \sqrt{\frac{1}{|\Omega|} \cdot 2^{2i+i/z} \cdot \gamma \cdot k \log k \cdot d_{\text{VC}} \cdot \log(k|\Omega|)} \cdot \mathbb{P}[\Omega \wedge \overline{\mathcal{E}}] \right)
\end{aligned}$$

Due to Corollary 4.5, event $\overline{E}$ happens with probability at most $\varepsilon \cdot 2^{-2i}/k \leq 2^{-2i}$. Therefore

$$\begin{aligned}
&\mathbb{E}_\Omega \mathbb{E}_g \sum_{h=1}^{h_{\max}-1} \left[\sup_{\mathcal{S}} \left| \frac{\frac{1}{|\Omega|} \sum_{C_j \in T_i} \sum_{p \in C_j \cap \Omega} (v_p^{\mathcal{S},h+1} - v_p^{\mathcal{S},h}) \cdot w_p \cdot g_p}{\text{cost}(G,\mathcal{S}) + \text{cost}(G,\mathcal{A})} \right|\right] \\
&\lesssim \left( \sqrt{\frac{1}{|\Omega|} \cdot \min\left(2^{i(1-1/z)}, k\right) \cdot \gamma \cdot k \log k \cdot d_{\text{VC}} \cdot \log(k|\Omega|)} + 2^{-2i} \cdot 2^{i+i/(2z)} \cdot \sqrt{\frac{1}{|\Omega|} \cdot \gamma \cdot k \log k \cdot d_{\text{VC}} \cdot \log(k|\Omega|)} \right) \\
&\lesssim \left( \sqrt{\frac{1}{|\Omega|} \cdot \min\left(\varepsilon^{-z+1}, k\right) \cdot \gamma \cdot k \log k \cdot d_{\text{VC}} \cdot \log(k|\Omega|)} + \sqrt{\frac{1}{|\Omega|} \cdot \gamma \cdot k \log k \cdot d_{\text{VC}} \cdot \log(k|\Omega|)} \right) \\
&\lesssim \frac{\varepsilon}{\log^4(k\varepsilon^{-1})}
\end{aligned}$$

where the second to last inequality holds by our choice of $\Omega$, $2^i \leq \varepsilon^{-z}$ and $x \geq y \log x$ being true for $y \log y \lesssim x$.

**Term 3** We will leverage the fact that the $q_p$ values are element-wise constant within each cluster by bounding the estimator via the expected maximum coreset error over all of the clusters in type $i$. To this end, notice that $T_i \subset G$ implies that we can write $\text{cost}(T_i, \mathcal{S}) \le \text{cost}(G, \mathcal{S}) \le \text{cost}(G, \mathcal{A}) + \text{cost}(G, \mathcal{S})$. This then gives:

$$\begin{aligned} & \mathbb{E}_\Omega \mathbb{E}_g \sup_{\mathcal{S}} \left[ \left| \frac{\sum_{C_j \in T_i} \sum_{p \in C_j \cap \Omega} q_p^{\mathcal{S}} \cdot w_p g_p}{|\Omega| \cdot (\text{cost}(G, \mathcal{S}) + \text{cost}(G, \mathcal{A}))} \right| \right] \\ \le \; & \mathbb{E}_\Omega \mathbb{E}_g \sup_{\mathcal{S}} \left[ \left| \frac{\sum_{C_j \in T_i} \sum_{p \in C_j \cap \Omega} q_p^{\mathcal{S}} \cdot w_p g_p}{|\Omega| \cdot \text{cost}(T_i, \mathcal{S})} \right| \right] \\ \le \; & \mathbb{E}_\Omega \mathbb{E}_g \sup_{\mathcal{S}} \left[ \left| \max_{C_j \in T_i} \frac{\sum_{p \in C_j \cap \Omega} q_p^{\mathcal{S}} \cdot w_p g_p}{|\Omega| \cdot \text{cost}(C_j, \mathcal{S})} \right| \right], \end{aligned}$$

where the last step is by the fact that $\sum \frac{a_j}{\sum b_j} \le \max \frac{a_j}{b_j}$ for positive numbers $a_j, b_j$. For every cluster, we have a net of size 1, as, up to scaling, $\frac{\sum_{p \in C_j \cap \Omega} q_p^{\mathcal{S}} \cdot w_p g_p}{|\Omega| \cdot \text{cost}(C_j, \mathcal{S})}$ is a single vector. Therefore, we have an overall net of size $k$.

Again, we condition on $\Omega$. We have $q_p^{\mathcal{S}} \cdot w_p = q_p^{\mathcal{S}} \cdot \frac{|C_i \cap G| \cdot \text{cost}(G, \mathcal{A})}{\text{cost}(C_i, \mathcal{A})} \lesssim 2^i \cdot \text{cost}(G, \mathcal{A}) \lesssim 2^i \cdot k \cdot \text{cost}(C_j, \mathcal{A}) \lesssim k \cdot \text{cost}(C_j, \mathcal{S})$. Therefore, the variance is

$$\frac{1}{|\Omega|^2} \frac{\sum_{p \in C_j \cap \Omega} \left(q_p^{\mathcal{S}} \cdot w_p\right)^2}{\text{cost}^2(C_j, \mathcal{S})} \lesssim \sum_{p \in C_j \cap \Omega} \frac{k^2}{|\Omega|}.$$

Conditioned on event $\mathcal{E}$ and Fact 4.6, this term is at most $\frac{k}{|\Omega|}$, otherwise, we bound it by $\frac{k^2}{|\Omega|}$ Again, using the law of total expectation and Lemma 4.2, we then obtain

$$\begin{aligned} & \mathbb{E}_\Omega \mathbb{E}_g \sup_{\mathcal{S}} \left[ \left| \frac{\sum_{C_j \in T_i} \sum_{p \in C_j \cap \Omega} q_p^{\mathcal{S}} \cdot w_p g_p}{|\Omega| \cdot (\text{cost}(G, \mathcal{S}) + \text{cost}(G, \mathcal{A}))} \right| \right] \\ \lesssim & \sqrt{\frac{k}{|\Omega|} \cdot \log k \cdot \mathbb{P}[\mathcal{E}]} + \sqrt{\frac{k^2}{|\Omega|} \cdot \log k \cdot \mathbb{P}[\overline{\mathcal{E}}]} \lesssim \sqrt{\frac{k}{|\Omega|} \cdot \log k} \end{aligned}$$

where the final inequality holds $\mathbb{P}[\overline{\mathcal{E}}] \le \varepsilon \cdot 2^{-2i}/k \le 1/k$ as per Corollary 4.5. Thus using the bounds on $|\Omega|$, we may conclude

$$\mathbb{E}_\Omega \mathbb{E}_g \sup_{\mathcal{S}} \left[ \left| \frac{\sum_{C_j \in T_i} \sum_{p \in C_j \cap \Omega} q_p^{\mathcal{S}} \cdot w_p g_p}{|\Omega| \cdot (\text{cost}(G, \mathcal{S}) + \text{cost}(G, \mathcal{A}))} \right| \right] \lesssim \frac{\varepsilon}{\log^4(k\varepsilon^{-1})}$$

as desired.

## 4.3 Proof of Lemma 2.5

The proof is identical to that of Lemma 15 from [21], and we write it here for completeness. While Lemma 15 applies to Eulcidean space, none of the properties of Euclidean space are used, and we likewise do not require properties of the VC dimension here. The idea is that the points in every huge cluster have the same cost up to a $(1 \pm \varepsilon)$ factor, so event $\mathcal{E}$ implies that the cost is well approximated.

Let us recall that we consider points $p \in C \cap G$ where $C \in T_{large}$, i.e., their cost in the current solution $\mathcal{S}$ is much larger than the cost they have in the approximate starting solution $\mathcal{A}$. By virtue of the point being so expensive, we have that its cost is roughly identical to the average cost of the entire cluster induced by $\mathcal{S}$ on $G$, so long as all induced cluster sizes are well-estimated. Our goal is to show for any $C \in T_{large}$

$$\mathbb{E}_\Omega \left[ \sup_{\mathcal{S}} \left| \frac{\frac{1}{|\Omega|} \sum_{p \in C \cap \Omega} w_p \text{cost}(p, \mathcal{S})}{\text{cost}(C, \mathcal{S})} - 1 \right| \right] \le \varepsilon, \tag{13}$$

By definition of $T_{large}$ and Lemma 3.2, we have for any two points $p, p' \in C$ $|\text{cost}(p,\mathcal{S}) - \text{cost}(p',\mathcal{S})| \lesssim \varepsilon \cdot \text{cost}(p,\mathcal{S}) + \varepsilon^{-z} \cdot \text{cost}(p,p') \lesssim \varepsilon \cdot \text{cost}(p,\mathcal{S})$. In other words, the cost of any two points is equal up to a $(1 \pm \varepsilon)$ factor. This implies that if event $\mathcal{E}$ holds, that is if $\frac{1}{|\Omega|}\sum_{p\in C\cap\Omega} w_p = (1\pm\varepsilon)\cdot|C|$, then $\frac{1}{|\Omega|}\sum_{p\in C\cap\Omega} w_p \text{cost}(p,\mathcal{S}) = (1\pm\varepsilon)^2\text{cost}(C,\mathcal{S})$.

If $\overline{\mathcal{E}}$, we have $0 \leq \frac{1}{\Omega}\sum_{p\in C\cap\Omega}\text{cost}(p,\mathcal{S})\cdot w_p \lesssim \frac{1}{\Omega}\sum_{p\in C\cap\Omega}\frac{\text{cost}(C,\mathcal{S})}{\text{cost}(C,\mathcal{A})}\cdot\text{cost}(G,\mathcal{A}) \lesssim k\cdot\text{cost}(C,\mathcal{S})$. Thus, by the law of total expectation

$$\mathbb{E}_\Omega\left[\frac{\frac{1}{|\Omega|}\sum_{p\in C\cap\Omega} w_p\text{cost}(p,\mathcal{S})}{\text{cost}(C,\mathcal{S})} - 1\right] \lesssim \varepsilon\cdot\mathbb{P}[\mathcal{E}] + k\cdot\mathbb{P}[\overline{\mathcal{E}}] \lesssim \varepsilon$$

where we use $\mathbb{P}[\overline{\mathcal{E}}] \leq \varepsilon/k$ from Corollary 4.5.

## 4.4 Proof of Lemma 2.6

The proof of this lemma is almost identical to the proof of Lemma 2.3, aside from a small changes when accounting for the sampling probability of a point.

We first apply the symmetrization, so we wish to show

$$\mathbb{E}_\Omega\mathbb{E}_g\left[\sup_{\mathcal{S}}\left|\frac{\frac{1}{|\Omega|}\sum_{C_j\in T_{close}}\sum_{p\in C_j\cap\Omega}\text{cost}(p,\mathcal{S})\cdot w_p\cdot g_p}{\text{cost}(G,\mathcal{S})+\text{cost}(G,\mathcal{A})}\right|\right] \leq \frac{\varepsilon}{\log^4(k\varepsilon^{-1})}.$$

We first write $\text{cost}(p,\mathcal{S})$ as a telescoping series. Specifically,

$$\text{cost}(p,\mathcal{S}) = \underbrace{v_p^{\mathcal{S}} - v_p^{\mathcal{S},h_{\max}}}_{\text{Term 1}} + \underbrace{\sum_{h=0}^{h_{\max}-1} v_p^{\mathcal{S},h+1} - v_p^{\mathcal{S},h}}_{\text{Term 2}},$$

where we determine $v_p^{\mathcal{S},h}$ as the value of $p$ drawn from a solution from the clustering net $\mathcal{N}_G(2^{-h},k,2^{2z})$ approximating $\mathcal{S}$, $2^{-h_{\max}} = \varepsilon^{2z+2}/k$, that is the maximum value of $\alpha$ in the clustering nets. We will control the Gaussian process for each term individually, that is, we write

$$\begin{aligned}
&\mathbb{E}_\Omega\mathbb{E}_g\left[\sup_{\mathcal{S}}\left|\frac{\frac{1}{|\Omega|}\sum_{C_j\in T_{close}}\sum_{p\in C_j\cap\Omega}\text{cost}(p,\mathcal{S})\cdot w_p\cdot g_p}{\text{cost}(G,\mathcal{S})+\text{cost}(G,\mathcal{A})}\right|\right]\\
\leq&\mathbb{E}_\Omega\mathbb{E}_g\left[\sup_{\mathcal{S}}\left|\frac{\frac{1}{|\Omega|}\sum_{C_j\in T_{close}}\sum_{p\in C_j\cap\Omega}(v_p^{\mathcal{S}} - v_p^{\mathcal{S},h_{max}})\cdot w_p\cdot g_p}{\text{cost}(G,\mathcal{S})+\text{cost}(G,\mathcal{A})}\right|\right]\\
&+\mathbb{E}_\Omega\mathbb{E}_g\sum_{h=1}^{h_{\max}-1}\left[\sup_{\mathcal{S}}\left|\frac{\frac{1}{|\Omega|}\sum_{C_j\in T_{close}}\sum_{p\in C_j\cap\Omega}(v_p^{\mathcal{S},h+1} - v_p^{\mathcal{S},h})\cdot w_p\cdot g_p}{\text{cost}(G,\mathcal{S})+\text{cost}(G,\mathcal{A})}\right|\right]
\end{aligned}$$

Showing that both terms are upper bounded by $\frac{\varepsilon}{\log^4(k\varepsilon^{-1})}$ yields the claim.

**Term 1** We apply the Cauchy Schwarz inequality to obtain

$$\begin{aligned}
&\mathbb{E}_\Omega\mathbb{E}_g\left[\sup_{\mathcal{S}}\left|\frac{\frac{1}{|\Omega|}\sum_{C_j\in T_{close}}\sum_{p\in C_j\cap\Omega}(v_p^{\mathcal{S}} - v_p^{\mathcal{S},h_{max}})\cdot w_p\cdot g_p}{\text{cost}(G,\mathcal{S})+\text{cost}(G,\mathcal{A})}\right|\right]\\
\leq&\frac{1}{|\Omega|}\mathbb{E}_\Omega\sqrt{\sum_{C_j\in T_{close}}\sum_{p\in C_j\cap\Omega}\left(\frac{(v_p^{\mathcal{S}} - v_p^{\mathcal{S},h_{max}})\cdot w_p}{\text{cost}(G,\mathcal{S})+\text{cost}(G,\mathcal{A})}\right)^2}\cdot\mathbb{E}_g\sqrt{\sum_{p\in C_j\cap\Omega} g_p^2}
\end{aligned}$$

The second factor may be controlled by $\sqrt{|\Omega|}$ due to Jensen's inequality and noting that $\mathbb{E}_g\sum_{p\in C_j\cap\Omega} g_p^2 = |C_j\cap\Omega|$. For the first term, we know that $|v_p^{\mathcal{S}} - v_p^{\mathcal{S},h_{max}}| \lesssim \varepsilon\cdot\varepsilon^{2z+2}/k\cdot v_p^{\mathcal{A}}$,since we are considering $T_{close}$.

Therefore

$$\sqrt{\sum_{C_j\in T_{close}}\sum_{p\in C_j\cap\Omega}\left(\frac{(v_p^{\mathcal{S}}-v_p^{\mathcal{S},h_{max}})\cdot w_p}{\text{cost}(G,\mathcal{S})+\text{cost}(G,\mathcal{A})}\right)^2}$$

$$\lesssim \varepsilon^{2z+2}/k\cdot\sqrt{\sum_{C_j\in T_{close}}\sum_{p\in C_j\cap\Omega}\left(\frac{v_p^{\mathcal{A}}\cdot\frac{\text{cost}(G,\mathcal{A})}{\text{cost}(p,\mathcal{A})}}{\text{cost}(G,\mathcal{S})+\text{cost}(G,\mathcal{A})}\right)^2}$$

$$\lesssim \varepsilon^{2z+2}/k\cdot\sqrt{\sum_{C_j\in T_{close}}\sum_{p\in C_j\cap\Omega}\left(\frac{\text{cost}(G,\mathcal{A})}{\text{cost}(G,\mathcal{S})+\text{cost}(G,\mathcal{A})}\right)^2}\lesssim \varepsilon^{2z+2}/k\cdot\sqrt{|\Omega|}$$

Hence, we may conclude that Term 1 is bounded by at most $\varepsilon^{2z+2}/k \lesssim \frac{\varepsilon}{\log^4(k\cdot\varepsilon^{-1})}$.

**Term 2** We condition on $\Omega$. Now $\frac{\frac{1}{|\Omega|}\sum_{C_j\in T_{close}}\sum_{p\in C_j\cap\Omega}(v_p^{\mathcal{S},h+1}-v_p^{\mathcal{S},h})\cdot w_p\cdot g_p}{\text{cost}(G,\mathcal{S})+\text{cost}(G,\mathcal{A})}$ is Gaussian distributed with mean 0 and variance

$$\frac{1}{|\Omega|^2}\frac{\sum_{C_j\in T_{close}}\left(\sum_{p\in C_j\cap\Omega}(v_p^{\mathcal{S},h+1}-v_p^{\mathcal{S},h})\cdot w_p\right)^2}{(\text{cost}(G,\mathcal{S})+\text{cost}(G,\mathcal{A}))^2}$$

$$\lesssim\frac{2^{-2h}}{|\Omega|^2}\frac{\sum_{C_j\in T_{close}}\sum_{p\in C_j\cap\Omega}\left(v_p^{\mathcal{S}}\cdot\frac{\text{cost}(G,\mathcal{A})}{\text{cost}(p,\mathcal{A})}\right)^2}{(\text{cost}(G,\mathcal{S})+\text{cost}(G,\mathcal{A}))^2}\lesssim\frac{2^{-2h}}{|\Omega|}.$$

Therefore, using Lemma 4.2, the bounds on the clustering nets for type $2z$ from Lemma 3.6 and the law of total expectation, we have

$$\mathbb{E}_\Omega\mathbb{E}_g\sum_{h=1}^{h_{\max}-1}\left[\sup_{\mathcal{S}}\left|\frac{\frac{1}{|\Omega|}\sum_{C_j\in T_{small}}\sum_{p\in C_j\cap\Omega}(v_p^{\mathcal{S},h+1}-v_p^{\mathcal{S},h})\cdot w_p\cdot g_p}{\text{cost}(G,\mathcal{S})+\text{cost}(G,\mathcal{A})}\right|\right]$$

$$\lesssim\sum_{\Omega}\mathbb{E}_g\sum_{h=1}^{h_{\max}-1}\left[\sup_{\mathcal{S}}\left|\frac{\frac{1}{|\Omega|}\sum_{C_j\in T_{small}}\sum_{p\in C_j\cap\Omega}2^{-h}v_p^{\mathcal{S}}\cdot w_p\cdot g_p}{\text{cost}(G,\mathcal{S})+\text{cost}(G,\mathcal{A})}\right|\Big|\Omega\right]\cdot\mathbb{P}[\Omega]$$

$$\lesssim\sum_{\Omega}\sum_{h=1}^{h_{\max}-1}\sqrt{\frac{2^{-2h}}{|\Omega|}\cdot\log\exp\left(\gamma 2^h\cdot k\log k\cdot d_{\text{VC}}\cdot\log|\Omega|\right)}\cdot\mathbb{P}[\Omega]$$

$$\lesssim\sum_{\Omega}\sum_{h=1}^{h_{\max}-1}\sqrt{\frac{2^{-h}}{|\Omega|}\cdot\gamma k\log k\cdot d_{\text{VC}}\cdot\log|\Omega|}\cdot\mathbb{P}[\Omega]\lesssim\frac{\varepsilon}{\log^4(k\varepsilon^{-1})}$$

where the final bound follows from our choice of $\Omega$ and $x\geq y\log x$ being true for $y\log y\lesssim x$.

## 4.5 Proof of Lemma 2.7

The proof is similar to that of Lemma 17 from [21]. The idea is that if points from an outer ring are served are served at more than a constant times their cost in $\mathcal{A}$, the entire set of points in the main rings becomes just as expensive. Since the number of points in the outer rings is very small compared those of the main rings, the contribution of the outer ring is now insignificant.

**Lemma 4.7.** *The following event $\mathcal{F}$ holds for all $C_j$*

- $\sum_{p\in(C_j\cap G)\cap\Omega}\text{cost}(p,\mathcal{A})\cdot w_p\lesssim k\cdot\text{cost}(C_j\cap G,\mathcal{A})$
- $\sum_{p\in(C_j\cap G)\cap\Omega}w_p\lesssim\frac{\varepsilon^{2z}}{k^2}|C_j|$

We note that unlike $\mathcal{E}$, $\mathcal{F}$ is a deterministic condition.

*Proof.* For the first claim, we note that $\text{cost}(p,\mathcal{A})\cdot w_p \leq k\text{cost}(G,\mathcal{A}) \lesssim k\cdot\text{cost}(C_j\cap G,\mathcal{A})$. Therefore $\frac{1}{|\Omega|}\sum_{p\in(Cj\cap G)\cap\Omega}\text{cost}(p,\mathcal{A})\cdot w_p \lesssim k\cdot\text{cost}(C_j\cap G,\mathcal{A})$.

For the second condition, we first recall that by definition of the outer rings, for any $p\in C_j\cap G$, we have $\text{cost}(p,\mathcal{A})\geq k^3\cdot\varepsilon^{-2z}\frac{\text{cost}(C_j,\mathcal{A})}{|C_j|}$. This implies

$$\begin{aligned}\frac{1}{|\Omega|}\sum_{p\in(C_j\cap G)\cap\Omega} w_p =& \frac{1}{|\Omega|}\sum_{p\in(C_j\cap G)\cap\Omega}\frac{\text{cost}(G,\mathcal{A})}{\text{cost}(p,\mathcal{A})}\\ \lesssim& \frac{1}{|\Omega|}\sum_{p\in(C_j\cap G)\cap\Omega}\frac{k\cdot\text{cost}(C_j\cap G,\mathcal{A})}{k^3\cdot\varepsilon^{-z}\frac{\text{cost}(C_j,\mathcal{A})}{|C_j|}} \lesssim \frac{\varepsilon^{2z}}{k^2}\cdot|C_j|.\end{aligned}$$

□

We can now conclude the claim. Let $C_j\in T_{far}$ and let $c_j$ be the center of $C_j$ in $\mathcal{A}$. We have for any point $p\in C_j$ due to Lemma 3.2

$$4^z\text{cost}(p,\mathcal{A})\leq\text{cost}(p,\mathcal{S})\leq 2^z\text{cost}(p,\mathcal{A})+2^z\text{cost}(c_j,\mathcal{S})$$

which implies $\text{cost}(c_j,\mathcal{S})\geq 2^z\text{cost}(p,\mathcal{A})\geq 2^z\varepsilon^{-2z}k^2\cdot\frac{\text{cost}(C_j,\mathcal{A})}{|C_j|}$. By Markov's inequality, there are at most $\frac{\varepsilon^{2z}}{k^2}\cdot|C_j|$ points that cost more than $\varepsilon^{-2z}k^2\cdot\frac{\text{cost}(C_j,\mathcal{A})}{|C_j|}$. Therefore

$$\text{cost}(C_j,\mathcal{S})\gtrsim\text{cost}(c_j,\mathcal{S})\cdot\left(1-\varepsilon^{-z}k^2\cdot\frac{\text{cost}(C_j,\mathcal{A})}{|C_j|}\right)\cdot|C_j|\gtrsim\varepsilon^{-2z}k^2\cdot\text{cost}(C_j,\mathcal{A}).$$

Using event $\mathcal{F}$, we now have

$$\begin{aligned}&\frac{1}{|\Omega|}\sum_{p\in(C_j\cap G)\cap\Omega} w_p\text{cost}(p,\mathcal{S})\\ \lesssim&\frac{1}{|\Omega|}\sum_{p\in(C_j\cap G)\cap\Omega} w_p(\text{cost}(p,\mathcal{A})+\text{cost}(c_j,\mathcal{S}))\\ \lesssim&\frac{1}{|\Omega|}\sum_{p\in(C_j\cap G)\cap\Omega} w_p\text{cost}(c_j,\mathcal{S}))+k\cdot\text{cost}(C_j\cap G,\mathcal{A})\\ \lesssim&\varepsilon^{2z}/k^2\cdot|C_j|\text{cost}(c_j,\mathcal{S}))+\frac{\varepsilon^2}{k}\cdot\text{cost}(C_j,\mathcal{S})\\ \lesssim&\frac{\varepsilon^2}{k}\cdot\text{cost}(C_j,\mathcal{S}).\end{aligned}$$

By a similar analysis, we can bound the cost of the $C_j\cap G$:

$$\begin{aligned}&\sum_{p\in C_j\cap G}\text{cost}(p,\mathcal{S})\\ \lesssim&\sum_{p\in C_j\cap G}(\text{cost}(p,\mathcal{A})+\text{cost}(c_j,\mathcal{S}))\\ \lesssim&\sum_{p\in C_j\cap G}\text{cost}(c_j,\mathcal{S}))+\text{cost}(C_j\cap G,\mathcal{A})\\ \lesssim&\varepsilon^{2z}/k^2\cdot|C_j|\text{cost}(c_j,\mathcal{S}))+\frac{\varepsilon^2}{k^2}\cdot\text{cost}(C_j,\mathcal{S})\\ \lesssim&\frac{\varepsilon^2}{k}\cdot\text{cost}(C_j,\mathcal{S}).\end{aligned}$$

Summing this up over all clusters in $T_{far}$ yields

$$\begin{aligned}
&\left|\sum_{C_j\in T_{far}}\left(\sum_{p\in(C_j\cap G)\cap\Omega}\text{cost}(p,\mathcal{S})\cdot w_p-\text{cost}(C_j\cap G,\mathcal{S})\right)\right| \\
&\le \sum_{C_j\in T_{far}}\left(\sum_{p\in(C_j\cap G)\cap\Omega}\text{cost}(p,\mathcal{S})\cdot w_p+\text{cost}(C_j\cap G,\mathcal{S})\right) \\
&\lesssim \sum_{C_j\in T_{far}}\frac{\varepsilon^2}{k}\text{cost}(C_j,\mathcal{S})\lesssim\frac{\varepsilon}{\log^4(k\varepsilon^{-1})}\text{cost}(P,\mathcal{S}),
\end{aligned}$$

as desired.

# 5 Applications

We briefly outline a few applications of Theorem 1.1. Similar bounds can also be given via Theorem 1.2 for other $(k,z)$ clustering objectives.

**Minor Free Graphs.** For $K_h$ minor free graphs, the VC dimension of the range space induced by metric balls with respect to shortest path distances is bounded by at most $h-1$, see [8, 17, 41]. This immediately gives rise to the following corollary.

**Corollary 5.1.** *Given a (potentially weighted) graph $G(V,E)$ excluding a minor of size $h$, there exists a coreset for $k$-median clustering with respect to shortest path distances of size*

$$\tilde{O}(kh\varepsilon^{-2}).$$

As mentioned in the introduction, for planar graphs in particular, this yields a coreset whose size is $\tilde{O}(k\varepsilon^{-2})$, which improves over the $\tilde{O}\left(k\varepsilon^{-6}\right)$ bound of Cohen-Addad et al. [18] and the $\tilde{O}\left(k^2\varepsilon^{-4}\right)$ bound by Braverman et al. [10]. More generally for minor free (but not necessarily planar) graphs, the previous results by Braverman et al. [10] yields a coreset of size $\tilde{O}\left(k^2\varepsilon^{-4}\cdot f(h)\right)$, where $f(h)$ is at least doubly exponential in $h$, and by Braverman et al. [11] yields a coreset of size $\tilde{O}\left(k^3h\varepsilon^{-3}\right)$.

**Polygonal Curves and Frechet Distance.** Another important set system are metric balls induced by the Frechet distance with respect to polygonal curves. A polygonal curve $P$ of length $m$ is a sequence of vertices $p_1,\dots p_m\in\mathbb{R}^d$, where consecutive pairs of vertices create edges. Given two curves $P_1$ and $P_2$, a warping path is a sequence $(1,1)=(i_1,j_1),(i_2,j_2),\dots,(i_M,j_M)=(|P_1|,|P_2|)$ such that $i_k-i_{k-1}$ and $j_k-j_{k-1}$ are either 0 or 1 for all $k$. The set of all warping paths between $P_1$ and $P_2$ are denoted by $\mathcal{W}_{P_1,P_2}$. Then the discrete Frechet distance is defined as

$$\text{dist}_{dF}(P_1,P_2)=\min_{w\in\mathcal{W}_{P_1,P_2}}\max_{(i,j)\in w}\|p_i-q_j\|,$$

where $\|x\|=\sqrt{\sum x_i^2}$ is the Euclidean norm.

In the continuous case, we view a polygonal curve $P:[0,1]\to\mathbb{R}^d$ by fixing $m$ values $0=t_1<\dots<t_m=1$ and defining $P(t)=\lambda p_{i+1}-(1-\lambda)p_i$ where $\lambda=\frac{t-t_i}{t_{i+1}t_i}$ for $t_i\le t\le t_{i+1}$. The continuous Frechet distance between two curves $P_1$ and $P_2$ is defined as

$$\text{dist}_F(P_1,P_2)=\inf_{\alpha,\beta:[0,1]\to[0,1]}\sup_{t\in[0,1]}\|P_1(\alpha(t))-P_2(\beta(t))\|.$$

In recent work, Brüning and Driemel [12] and Cheng and Huang [16] showed that if the center curves have length at most $\ell$ and the input curves have length at most $m$, the VC dimension of metric balls for both discrete and continuous Frechet distance may be bounded by $O(d\ell\log(\ell m))$, see also [26] for earlier, slightly weaker bounds. The $(k,\ell)$ clustering problem with respect to the Frechet distance uses at most $k$ centers of length at most $\ell$, while minimizing the sum of Frechet distances. For this problem, we obtain the following result:

**Corollary 5.2.** *Given a set of polygonal curves of length at most $m$, there exists a coreset for $k$ median clustering with respect to the discrete or continuous Frechet distance of size*

$$\tilde{O}(kd\ell\log(m)\varepsilon^{-2}).$$

The previous state of the art seems to be a recent result Conradi et al. [24], giving coresets of size $\tilde{O}\left(k^2 d\ell\varepsilon^{-2}\log m\log n\right)$, or alternatively trading the dependency on $\log n$ with a dependency on $k/\varepsilon$, the $\tilde{O}\left(k^3 d\ell\varepsilon^{-3}\log m\right)$ result by Braverman et al. [11].

**Hausdorff Metric.** Given two sets $X, Y \subseteq \mathbb{R}^d$, the directed Hausdorff distance from $X$ to $Y$ is defined as

$$\text{dist}_{\overrightarrow{H}}(X,Y) = \sup_{x\in X}\inf_{y\in Y}\|x-y\|$$

and the Hausdorff distance between $X$ and $Y$ is defined as

$$\text{dist}_H(X,Y) = \max\left(\text{dist}_{\overrightarrow{H}}(X,Y), \text{dist}_{\overrightarrow{H}}(Y,X)\right).$$

We consider polygonal regions, that is the border of a set $X$ is a polygonal curve with $p_1 = p_m$. The $(k,\ell)$-clustering problem asks to minimize the distance to the closest region with respect to the Hausdorff distance of at most $k$ candidate center regions where the border of each center region is at most $\ell$. For this metric, Brüning and Driemel [12] showed that the VC dimension is of the order $O(d\ell\log(\ell m))$, which combined with Theorem 1.1 implies

**Corollary 5.3.** *Given a set of polygonal regions with boundary length at most $m$, there exists a coreset for $(k,\ell)$ clustering with respect to the Hausdorff distance of size*

$$\tilde{O}(kd\ell\log(m)\varepsilon^{-2}).$$

While it is likely that results by [13, 24] would also yield coresets for this problem, the only stated results we are aware of are due to Braverman et al. [11] who gave a coreset of size $\tilde{O}\left(k^3\varepsilon^{-3}d^2\ell^2\log m\right)$. The stated result was obtained via the looser VC-dimension bound by Driemel et al. [26] and using the improved bounds by Brüning and Driemel [12], their coreset size should improve to $\tilde{O}\left(k^3\varepsilon^{-3}d\ell\log m\right)$.

# 6 A Brief Remark On An Error in a Previous Version

We had previously claimed a coreset bound of $\tilde{O}\left(k\cdot d_{VC}\cdot\left(\varepsilon^{-2}+\varepsilon^{-z}\right)\right)$ for general $(k,z)$ clustering. For $k$-median clustering, the bounds are identical, but Theorem 1.2 is slightly weaker. We believe that the aforementioned bound is not obtainable and conjecture that the bounds given in Theorem 1.2 are optimal for a worst-case metric for $\varepsilon^{-1} \ll k$, noting that there exist metrics where a VC-dimension based analysis does not yield an optimal coreset size.

To prevent similar errors, we briefly highlight some misleading and wrong claims we encountered in literature. Given two functions $\mathcal{F}$ and $\mathcal{G}$ with set systems induced by $R_f(i,t) = \{x|f_i(x)\geq t\}$ and $R_g(i,t) = \{x|g_i(x)\geq t\}$, for $f_i\in\mathcal{F}$ and $g_i\in\mathcal{G}$, suppose that the VC dimensions are $d_F$ and $d_G$ respectively. The VC dimension of $\mathcal{F}-\mathcal{G}$ with ranges $R(i,j,t) = \{x|f_i(x)-g_j(x)\geq t\}$ does not have bounded VC dimension, even if $d_F$ and $d_G$ are 1. To see this, suppose $\mathcal{F}$ and $\mathcal{G}$ are defined on the positive integers. Let $\mathcal{A}$ be the set of all functions mapping the positive integers to $\{0,1\}$ and let $\mathcal{F}$ consists of the functions of $2x + a_i(x)$, where $a_i\in\mathcal{A}$. Let $\mathcal{G}$ consist only of the function $2x$. Since $\mathcal{F}$ and $\mathcal{G}$ act monotonously on the positive integers, they have VC dimension 1. However, $\mathcal{F}-\mathcal{G}$ yields $\mathcal{A}$, which contains the indicator function for every set of positive integers and thus can shatter any set of positive integers.

## 7 Funding Acknowledgements

Andrew Draganov and Chris Schwiegelshohn are partially supported by the Independent Research Fund Denmark (DFF) under a Sapere Aude Research Leader grant No 1051-00106B. Matteo Russo is partially supported by the ERC Advanced Grant 788893 AMDROMA "Algorithmic and Mechanism Design Research in Online Markets", by the FAIR (Future Artificial Intelligence Research) project PE0000013, funded by the NextGenerationEU program within the PNRR-PE-AI scheme (M4C2, investment 1.3, line on Artificial Intelligence), by the PNRR MUR project IR0000013-SoBigData.it, and by the MUR PRIN grant 2022EKNE5K (Learning in Markets and Society).

# A Omitted Proofs

## A.1 Proof of Lemma 3.3

**Lemma 3.3.** *Let $\mathcal{A}$ be an approximate solution, $C$ be any cluster from this solution, and $G$ be a layered group obtained from $\mathcal{A}$. Let $p$ be any point in cluster $C \cap G$. Furthermore, assume $C$ is in type $i$ with respect to candidate solution $\mathcal{S}$. Then for some sufficiently large constant $\gamma_1$, we have*

$$\text{cost}(p, \mathcal{S}) \leq \gamma_1 \cdot (2^{i/z} \text{dist}^{z-1}(p\mathcal{S}) \text{dist}(p, \mathcal{A}) + \text{dist}^z(p, \mathcal{A})).$$

*Proof.* By definition of type $i$, there exists a point $p'$ in $C$ that satisfies $\text{cost}(p', \mathcal{S}) \leq 2^{i+1} \text{cost}(p', \mathcal{A})$. Then repeatedly using Lemma 3.2 and the definition of type $i$ and the group properties

$$\begin{aligned}
\text{cost}(p, \mathcal{S}) &\lesssim \text{cost}(p', \mathcal{S}) + \text{cost}(p, p') \\
&\lesssim 2^i \text{cost}(p', \mathcal{A}) + \text{cost}(p, \mathcal{A}) + \text{cost}(p', \mathcal{A}) \\
&\lesssim 2^i \text{cost}(p, \mathcal{A}) + \text{cost}(p, \mathcal{A}) \\
&\lesssim 2^{i/z} \cdot \left(2^{i/z} \text{dist}(p, \mathcal{A})\right)^{z-1} \cdot \text{dist}(p, \mathcal{A}) + \text{cost}(p, \mathcal{A}) \\
&\lesssim 2^{i/z} \cdot \text{dist}^{z-1}(p, \mathcal{S}) \cdot \text{dist}(p, \mathcal{A}) + \text{cost}(p, \mathcal{A})
\end{aligned}$$

□

## A.2 Proof of Lemma 3.4

**Lemma 3.4.** *Let $\mathcal{A}$ be an approximate solution, $C$ be any cluster from this solution, and $G$ be a layered group obtained from $\mathcal{A}$. Let $p$ be any point in cluster $C \cap G$. Furthermore, assume $C$ is in type $i$ with respect to candidate solution $\mathcal{S}$. Assume also that $2^z < \gamma_1$ for some constant $\gamma_1$. Then*

$$\text{cost}(p, \mathcal{S}) - q_p \leq (2 + \gamma_1) \cdot 2^{i(1-1/z)} \cdot \text{cost}(p, \mathcal{A}) \leq (2 + \gamma_1) \cdot \text{dist}(p, \mathcal{S})^{z-1} \cdot \text{dist}(p, \mathcal{A}).$$

*Proof.* We start by writing out $\text{cost}(p, \mathcal{S}) - q_p$ in terms of the triangle inequality for powers. Let $q(p) = \arg\min_{p' \in C} \text{cost}(p', \mathcal{S})$ be the point which induces $q_p$:

$$\text{cost}(p, \mathcal{S}) - q_p \leq \beta \cdot \text{dist}^z(p, \mathcal{S}) + \left(\frac{2z + \beta}{\beta}\right)^{z-1} \text{dist}^z(p, q(p)). \tag{14}$$

We will simplify Equation (14) using the following inequalities:

1. We have
$$\left(\frac{2z + \beta}{\beta}\right)^{z-1} \leq \left(\frac{2z}{\beta} + 1\right)^{z-1} \leq \frac{(2z)^z}{\beta^{z-1}}.$$

2. Because $p$ and $q_p$ are in the same cluster from the same group, we have
$$\begin{aligned}
\text{cost}(p, q(p)) &= \text{dist}^z(p, q(p)) \\
&\leq (\text{dist}(p, \mathcal{A}) + \text{dist}(q(p), \mathcal{A}))^z && \text{(Triangle Inequality)} \\
&\leq (3 \cdot \text{dist}(p, \mathcal{A}))^z && \text{(Group Definition)} \\
&\leq 2^{2z} \text{dist}^z(p, \mathcal{A})
\end{aligned}$$

3. By cluster $C$ belonging to type $i$, we have $\text{cost}(p, \mathcal{S}) \leq 2^{i+1} \text{cost}(p, \mathcal{A})$.

We now set $\beta = 2^{-i/z}$ and plug the above inequalities into Equation (14):

$$\text{cost}(p, \mathcal{S}) - \text{cost}(q(p), \mathcal{S}) \leq \beta \cdot \text{cost}(p, \mathcal{S}) + \left(\frac{2z + \beta}{\beta}\right)^{z-1} \text{cost}(p, q(p))$$

$$
\begin{aligned}
&\leq \beta \cdot 2 \cdot 2^{i}\mathrm{cost}(p,\mathcal{A}) + \beta^{-(z-1)} \cdot 2^{2z} \cdot (2z)^{z}\mathrm{cost}(p,\mathcal{A}) \\
&\leq 2 \cdot 2^{i(1-\frac{1}{z})} \cdot \mathrm{cost}(p,\mathcal{A}) + 2^{\frac{i}{z}(z-1)} \cdot 2^{2z} \cdot (2z)^{z} \cdot \mathrm{cost}(p,\mathcal{A}) \\
&= \left(2 \cdot 2^{i(1-\frac{1}{z})} + 2^{2z} \cdot 2^{\frac{i}{z}(z-1)}\right) \mathrm{dist}^{z}(p,\mathcal{A}) \qquad (15) \\
&= (2+\gamma_1^2) \cdot 2^{i(1-\frac{1}{z})} \cdot \mathrm{cost}(p,\mathcal{A}).
\end{aligned}
$$

This completes the first inequality of the lemma statement. We now show the second inequality:

$$
\begin{aligned}
(2+\gamma_1^2) \cdot 2^{i(1-\frac{1}{z})} \cdot \mathrm{cost}(p,\mathcal{A}) &= (2+\gamma_1^2) \cdot 2^{i(1-\frac{1}{z})} \cdot \mathrm{dist}^{z(1-\frac{1}{z})}(p,\mathcal{A}) \cdot \mathrm{dist}(p,\mathcal{A}) \\
&= (2+\gamma_1^2) \cdot \left(2^{i}\mathrm{dist}^{z}(p,\mathcal{A})\right)^{1-\frac{1}{z}} \cdot \mathrm{dist}(p,\mathcal{A}) \\
&\leq (2+\gamma_1^2) \cdot \mathrm{dist}^{z-1}(p,\mathcal{S}) \cdot \mathrm{dist}(p,\mathcal{A}),
\end{aligned}
$$

where the last inequality is by definition of the types. This completes the proof, since we assumed $\gamma_1$ to be a constant. □